\documentclass[pra,reprint,bibnotes,floatfix]{revtex4-2}  

\usepackage[T1]{fontenc}
\usepackage[utf8]{inputenc} 
\usepackage[australian]{babel}
\usepackage[a4paper,centering,hmargin=1.75cm,vmargin=2cm]{geometry} 
\usepackage{amsmath,amssymb,graphicx,bm,microtype} 
\usepackage[dvipsnames]{xcolor}
\usepackage[colorlinks,allcolors=blue!50!black]{hyperref}
\usepackage[all]{hypcap} 
\usepackage{booktabs,braket,cleveref,enumitem,titlesec,siunitx}
\titleformat{\paragraph}[runin]{\bfseries}{}{0pt}{}[.]
\newcommand{\abs}[1]{\lvert #1 \rvert}
\renewcommand{\vec}[1]{\bm{\mathrm{#1}}}

\begin{document}

\title{Delocalisation enables efficient charge generation in organic photovoltaics, even with little to no energetic offset}

\author{Daniel Balzer}
\affiliation{School of Chemistry and University of Sydney Nano Institute, University of Sydney, NSW 2006, Australia}

\author{Ivan Kassal}
\email[Email: ]{ivan.kassal@sydney.edu.au}
\affiliation{School of Chemistry and University of Sydney Nano Institute, University of Sydney, NSW 2006, Australia}

\begin{abstract}
Organic photovoltaics (OPVs) are promising candidates for solar-energy conversion, with device efficiencies continuing to increase. However, the precise mechanism of how charges separate in OPVs is not well understood because low dielectric constants produce a strong attraction between the charges, which they must overcome to separate. Separation has been thought to require energetic offsets at donor-acceptor interfaces, but recent materials have enabled efficient charge generation with small offsets, or with none at all in neat materials. Here, we extend delocalised kinetic Monte Carlo (dKMC) to develop a three-dimensional model of charge generation that includes disorder, delocalisation, and polaron formation in every step from photoexcitation to charge separation. Our simulations show that delocalisation dramatically increases charge-generation efficiency, partly by enabling excitons to dissociate in the bulk. Therefore, charge generation can be efficient even in devices with little to no energetic offset, including neat materials. Our findings demonstrate that the underlying quantum-mechanical effect that improves the charge-separation kinetics is faster and longer-distance hops between delocalised states, mediated by hybridised states of exciton and charge-transfer character.
\end{abstract}

\maketitle

In organic photovoltaics (OPVs), understanding how charges overcome their large Coulomb attraction to separate efficiently has remained elusive~\cite{Clarke2010,Few2015,Hou2018}. Due to the low dielectric constants in organic semiconductors ($\varepsilon_r\approx3-4$), photoexcitation produces tightly bound excitons. To overcome the large exciton-binding energies and enable charges to separate, OPVs are made by blending donor and acceptor materials to form a heterojunction. At the donor-acceptor interface, the offset between the energy levels of the two materials produces a driving force to dissociate the exciton, resulting in one of the charges moving across the interface to form a slightly separated charge-transfer (CT) state. Nevertheless, charges in a CT state are still Coulombically bound to each other with a strength more than an order of magnitude larger than the available thermal energy, an obstacle they must overcome to separate. The efficiency of the separation process is the internal quantum efficiency (IQE), the proportion of excitons that generate separated charges.

The near-unity IQE in some OPVs has been attributed to delocalisation, with several groups observing efficient and fast separation of delocalised charges~\cite{Bakulin2012,Jailaubekov2013,Grancini2013,Gelinas2014,Falke2014,Tamai2017}. The delocalisation enhancement has been attributed to a reduced electron-hole binding energy when the charges in a CT state delocalise and their separation increases. However, charge separation is a multi-step process, and when disorder and entropy are considered, the free-energy barrier to separation disappears~\cite{Hood2016} and actually increases when delocalisation is included~\cite{Gluchowski2018}, indicating that delocalisation enhancements must come from non-equilibrium kinetic effects, rather than solely energetic ones~\cite{Shi2017,Gluchowski2018}. Therefore, a full kinetic model is required for studying the effects of delocalisation on charge separation. Significant progress has been made towards this goal, with many kinetic models including delocalisation and finding improved efficiencies~\cite{Tamura2013,Bittner2014,Sun2014,Smith2014,Smith2015,Abramavicius2016,DAvino2016,Jankovic2017,Jankovic2018,Yan2018,Jankovic2020,Peng2022}. However, the computational difficulty of tracking the correlated motion of two quantum-mechanically delocalised charges limits these simulations to small systems or reduced dimensionality. Recently, we overcame this challenge to produce three-dimensional simulations of charge separation, finding that delocalisation enhancements to CT-separation efficiency are indeed a kinetic effect, with improved IQE even though delocalisation increases the binding energy~\cite{Balzer2022}.

\begin{figure*}
    \centering
    \vspace{-3mm}
    \includegraphics[width=0.9\textwidth]{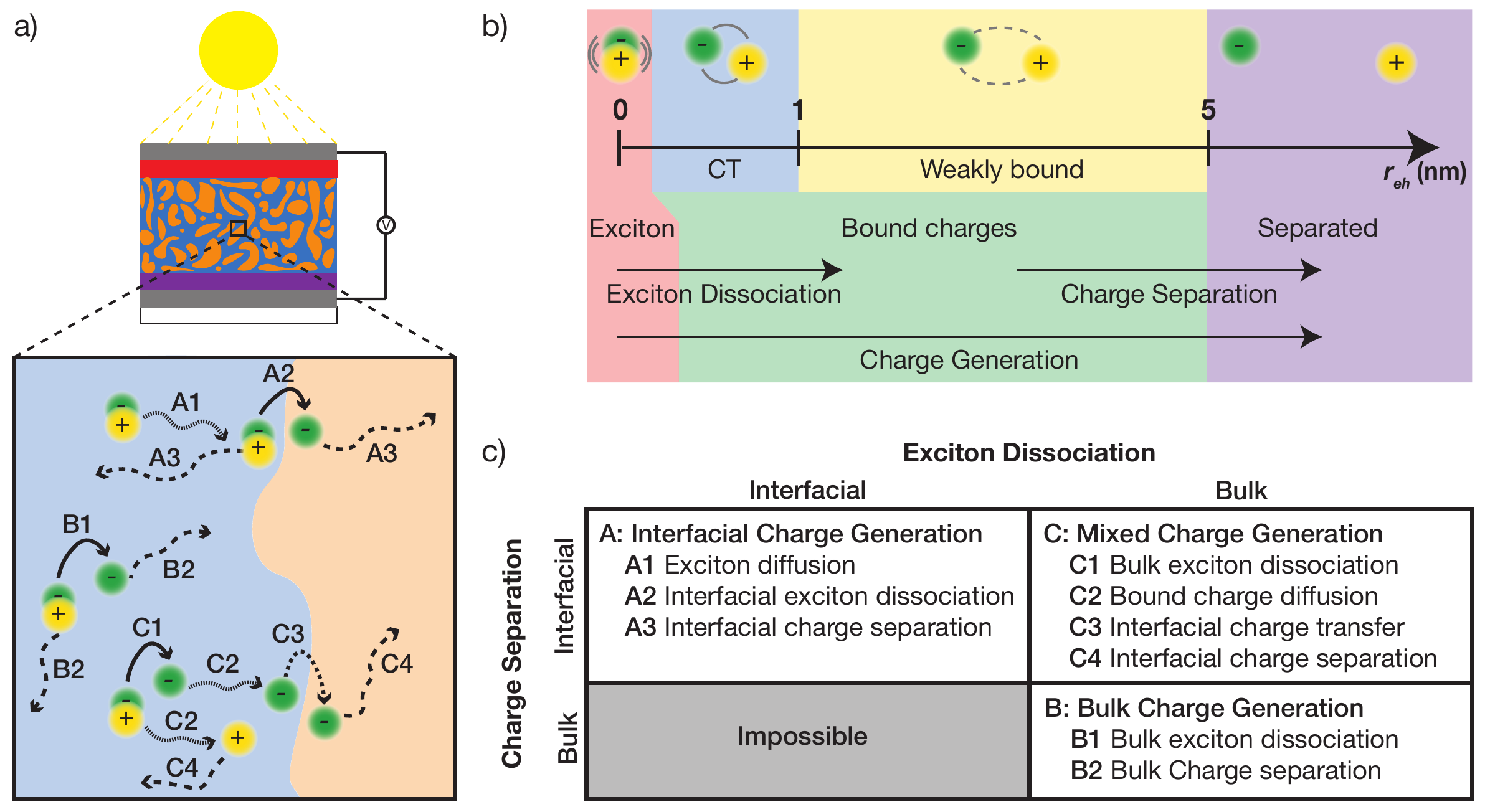}
    \caption{\textbf{Pathways of delocalised charge generation.} 
    \textbf{a)}~Charges in organic solar cells are usually assumed to separate via pathway A, interfacial charge generation: (A1) exciton diffuses to an interface where (A2) an energetic offset drives exciton dissociation into an interfacial CT state out of which (A3) charges separate. When excitons and charges become delocalised, they can separate more efficiently through pathway A, but also without an interfacial energetic offset via pathway B, bulk charge generation, where (B1) exciton dissociates and (B2) the charges separate within the single material. Delocalisation also enables pathway C, mixed charge generation: (C1) exciton dissociates in the bulk, before (C2) the bound charges diffuse to the interface, where (C3) charge transfer occurs across and (C4) the charges can separate in separate domains. 
    \textbf{b)}~Classification of the types of site-pairs and the processes between them. 
    \textbf{c)}~Classification of the charge generation pathways.}
    \label{fig:cartoon_mechanism}
\end{figure*}

Recent developments have further challenged conventional assumptions about the mechanism of charge generation. In particular, non-fullerene acceptors (NFAs) have produced efficient OPV devices~\cite{Li_Chao_2021,Zhu2021,Zheng2022} without the large energetic offsets thought to be required for exciton dissociation~\cite{Bin2016,Hou2018,Perdigon2020}. Explaining this observation is an outstanding theoretical challenge~\cite{Hou2018}, with one hypothesis being that low offsets enable the formation of hybridised states with the character of both excitons and separated charges~\cite{Qian2018,Eisner2019,Coropceanu2019,Qian2023}. Furthermore, experimental evidence suggests that delocalised and partially separated intermediate states form within NFA phases before reaching an interface~\cite{Wang2020,Tu2020,Zhu2021,Zhang2020,Dimitriev2022,Li2023,Li2023_2}, suggesting separation may not proceed through the typical pathway involving interfacial CT states. An even greater theoretical challenge is explaining the detection of free carriers in devices made of neat materials, where there is no interfacial energetic offset, including Y6~\cite{Price2022,Zhang2022,Saglamkaya2023}.

Here, we show that delocalisation can enable efficient charge generation in OPVs with little or no energetic offset. To do so, we extend our delocalised kinetic Monte Carlo (dKMC) algorithm---which has already been used to show that delocalisation improves charge and exciton transport~\cite{Balzer2020, Balzer2023} and the separation of charges from a CT state~\cite{Balzer2022}---to consider the full charge generation process, from excitons to separated charges. dKMC includes all of the essential ingredients: disorder, delocalisation, and polaron formation. We find that delocalisation produces large charge-generation improvements, especially in systems with little to no energetic offset, helping to explain the efficient low-offset and homojunction devices. 
The mechanism of the improved exciton dissociation is a kinetic, quantum effect: delocalised hops of greater speed and distance, mediated by hybridised states, allow separation to out-compete recombination. Overall, our findings support both of the hypotheses above, that delocalisation enables excitons to fully or partially separate in the bulk of one phase and that charge generation involves hybridisation of exciton and weakly bound states.

\begin{figure*}
    \centering
    \vspace{-3mm}
    \includegraphics[width=0.9\textwidth]{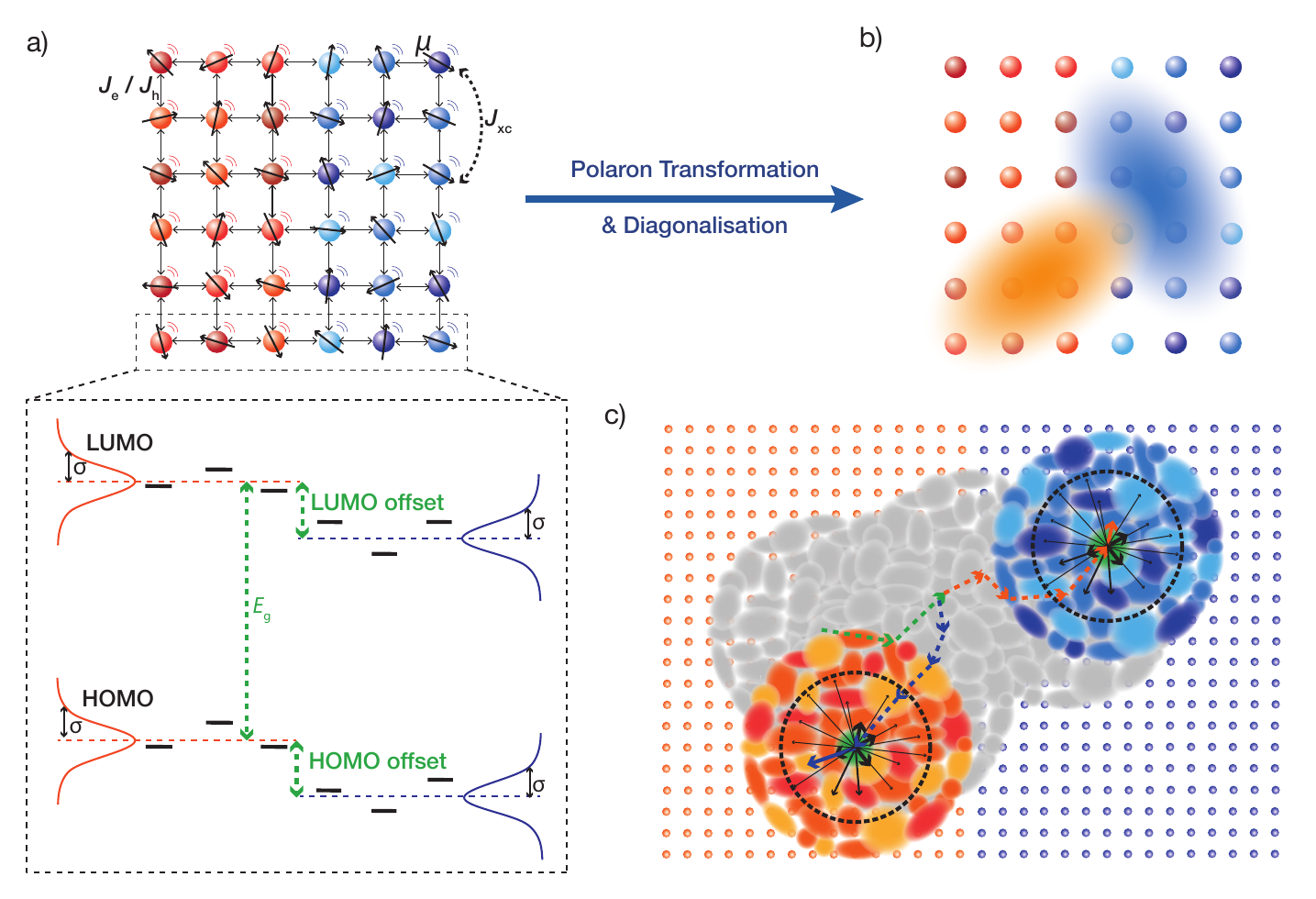}
    \vspace{-3mm}
    \caption{\textbf{dKMC model of charge separation.} \textbf{a)} A planar heterojunction is represented by a cubic lattice of donor (orange) and acceptor (blue) sites. Each site is coupled to its environment (motion lines) and is assigned a randomly oriented transition dipole moment $\vec{\mu}$ and disordered HOMO and LUMO energies (different shades, see inset).  
    There are nearest-neighbour electronic couplings $J_\mathrm{e}$ and $J_\mathrm{h}$ for electrons and holes and long-range exciton couplings $J_\mathrm{xc}$.
    \textbf{b)}~Each eigenstate of the polaron-transformed system Hamiltonian describes the simultaneous wavefunction of the delocalised electron (blue) and hole (orange). The cartoon depicts the charges as separable, but they are correlated in reality. 
    \textbf{c)}~Dynamics though the polaron states is propagated using dKMC, which tracks joint trajectories of both charges. After each hop, polaron states are re-calculated within the neighbourhood of the charges, outgoing rates (black arrows) are calculated to states within a cutoff distance (black circles), and the hopping destination is chosen probabilistically based on these outgoing rates. 
    }
    \label{fig:model}
\end{figure*}

\section{Methods\label{sec:model}}

We begin by describing the dKMC model of charge generation. The details of the underlying quantum master equation, the approximations, and the full algorithm are in secs.~S1--S5 of the Supporting Information (SI).

\subsection{Model}

We model an OPV heterojunction using a tight-binding model on a cubic lattice with lattice spacing $a$ and $N^d$ sites (molecules or parts of molecules), where $N$ is the number of sites along each of $d$ dimensions. The left $N^d/2$ sites represent the donor and the rest the acceptor~(\cref{fig:model}a). Every site is assigned a randomly oriented transition dipole moment $\vec{\mu}_n$ with constant magnitude $\mu$ as well as HOMO and LUMO energies for a hole or an electron occupying that site, respectively.

Energetic disorder is modelled by drawing site energies from Gaussian distributions $\mathcal{N}(E_0,\sigma)$, centred on $E_0$ and with standard deviation $\sigma$ measuring the disorder~\cite{Bassler1993}. There are four distributions (\cref{fig:model}): the donor HOMO energy centred at $0$, the donor LUMO centred at $E_g$, the acceptor HOMO centred at $-E^\mathrm{HOMO}_\mathrm{offset}$, and the acceptor LUMO centred at $E_g-E^\mathrm{LUMO}_\mathrm{offset}$.

The energies drawn from these four distributions must be correlated to ensure that the exciton disorder is smaller than the electronic disorder. For uncorrelated LUMO and HOMO energies with disorders $\sigma_\mathrm{LUMO}$ and $\sigma_\mathrm{HOMO}$ and a constant exciton-binding energy $E_b$, the disorder in the exciton energy $E_\mathrm{xc}=E^\mathrm{LUMO}-E^\mathrm{HOMO}-E_b$ would be $\sigma_\mathrm{xc}=\sqrt{\sigma_\mathrm{e}^2+\sigma_\mathrm{h}^2}$, which is larger than either electronic disorder. However, $\sigma_\mathrm{xc}$ should be less than $\sigma_\mathrm{e}$ and $\sigma_\mathrm{h}$ because local electrostatic variations shift HOMO and LUMO energies in the same direction (\cref{fig:model}a).
Therefore, we correlate the HOMO and LUMO energies to achieve a correlation $\rho = \mathrm{Cov}(E^\mathrm{LUMO},E^\mathrm{HOMO})/\sigma_\mathrm{HOMO}\sigma_\mathrm{LUMO}$, which gives  $\sigma^2_\mathrm{xc}=\sigma^2_\mathrm{LUMO}+\sigma^2_\mathrm{HOMO}-2\mathrm{Cov}(E^\mathrm{LUMO},E^\mathrm{HOMO})$. 
Assuming for simplicity that $\sigma=\sigma_\mathrm{LUMO}=\sigma_\mathrm{HOMO}$ leads to
$\rho = 1-\sigma^2_\mathrm{xc}/2\sigma^2$, meaning that $\sigma_\mathrm{xc}$ can be chosen independently of $\sigma$. The correlated energies for the donor are drawn from the bivariate normal distribution
\begin{equation}
\label{eq:energy_distribution}
\begin{bmatrix}
E_n^\mathrm{HOMO} \\ 
E_n^\mathrm{LUMO}
\end{bmatrix}
\sim \mathcal{N}\left(
\begin{bmatrix}
0 \\ 
E_g
\end{bmatrix},
\begin{bmatrix}
\sigma^2 & \rho\sigma^2 \\ 
\rho\sigma^2 & \sigma^2
\end{bmatrix}
\right),
\end{equation}
while those for the acceptor are drawn in the same way, but shifted by $[-E^\mathrm{HOMO}_\mathrm{offset}, -E^\mathrm{LUMO}_\mathrm{offset}]$. Throughout this work, we assume $\sigma=\SI{150}{meV}$ and $\sigma_\mathrm{xc}=\SI{30}{meV}$, corresponding to $\rho=0.98$.

As exciton dissociation is a two-particle problem, the Hilbert space is spanned by site-pairs $\ket{m,n}$, which describe an electron on site $m$ and a hole on site $n$. In this basis, the system is described by the Hamiltonian
\begin{multline}
    H_\mathrm{S} = \sum_{m,n} E_{mn} \ket{m,n}\bra{m,n} + \sum_{m\neq m', n} J_\mathrm{e} \ket{m,n}\bra{m',n} \\ + \sum_{m, n\neq n'} J_\mathrm{h} \ket{m,n}\bra{m,n'} + \sum_{o \neq o'} J_\mathrm{xc} \ket{o,o}\bra{o',o'}
    \label{eq:H_S},
\end{multline}
where the energy of a site-pair with electron-hole separation $r_{mn}$ is $E_{mn}=E_m^\mathrm{LUMO}-E_n^\mathrm{LUMO} - U(r_{mn})$, including the Coulomb potential of
\begin{equation}
    U(r_{mn})=
    \begin{cases}
    -\frac{e^2}{4\pi\varepsilon_0 \varepsilon_r r_{mn}} & \text{if } r_{mn} \neq 0\\
    E_b & \text{if } r_{mn} = 0
\end{cases},
\end{equation}
where $e$ is the elementary charge, $\varepsilon_0$ the vacuum permittivity, and $\varepsilon_r$ the dielectric constant (here, $\varepsilon_r=3.5$).

\Cref{eq:H_S} shows that site-pair $\ket{m,n}$ is coupled to other site-pairs $\ket{m',n'}$ in three cases. 
First, if only the electron is moving ($n=n'$), the two site-pairs are coupled by the electron coupling $J_\mathrm{e}$. 
Second, if only the hole is moving ($m=m'$), they are coupled by the hole coupling $J_\mathrm{h}$. We assume constant, nearest-neighbour $J_\mathrm{e}$ and $J_\mathrm{h}$, but this assumption could easily be relaxed to allow for disordered or non-nearest-neighbour couplings.  
Third, if both site-pairs represent excitons ($m=n$ and $m'=n'$), they are coupled by the long-range dipole-dipole coupling $J_\mathrm{xc}=\xi_{mm'}\mu^2/4\pi \varepsilon_0 \abs{\vec{r}_{mm'}}^3$, where the distance vector between $m$ and $m'$ is $\vec{r}_{mm'}$ and the orientation factor is $\xi_{mm'}=\hat{\vec{\mu}}_m\cdot\hat{\vec{\mu}}_{m'} - 3( \hat{\vec{r}}_{mm'}\cdot\hat{\vec{\mu}}_m)(\hat{\vec{r}}_{mm'}\cdot\hat{\vec{\mu}}_{m'})$, with hats indicating unit vectors. 

The environment is modelled as an independent bath of harmonic oscillators on each site~\cite{kohler2015textbook,MayKuhn}, 
\begin{equation}
     \label{eq:H_B}
     H_\mathrm{B}=\sum_{m,k}\omega_{mk} b^\dag_{mk}b_{mk},
\end{equation}
where $\omega_{nk}$ is the frequency of the $k$th mode on site $m$, with creation and annihilation operators $b^\dag_{nk}$ and $b_{nk}$.

The interaction between the system and the environment is modelled as a linear coupling of each site to each of its bath modes, with separate coupling strengths for electrons $g^\mathrm{e}_{mk}$, holes $g^\mathrm{h}_{nk}$, and excitons $g^\mathrm{xc}_{ok}$,
\begin{multline}
     \label{eq:H_SB}
     H_\mathrm{SB} = \sum_{m,n\neq m,k} g^\mathrm{e}_{mk}\ket{m,n}\bra{m,n}(b^\dag_{mk} + b_{mk})\\
    + \sum_{m,n\neq m,k} g^\mathrm{h}_{nk}\ket{m,n}\bra{m,n}(b^\dag_{nk} + b_{nk})\\
    +\sum_{o,k} g^\mathrm{xc}_{ok}\ket{o,o}\bra{o,o}(b^\dag_{ok} + b_{ok}).
\end{multline}

Crucial to OPVs is the formation of polarons, quasi-particles consisting of charges or excitons dressed by the induced distortion in the environment~\cite{Frolich1954,Holstein1959}. Polarons are incorporated using the polaron transformation~\cite{Grover1971} 
\begin{multline}
     e^S = \exp\Bigg(\sum_{m,n\neq m,k} \frac{g^\mathrm{e}_{mk}}{\omega_{mk}}\ket{m,n}\bra{m,n}(b^\dag_{mk}-b_{mk}) \\
     + \sum_{m,n\neq m,k}\frac{g^\mathrm{h}_{nk}}{\omega_{nk}}\ket{m,n}\bra{m,n}(b^\dag_{nk}-b_{nk}) \\
     +
     \sum_{o,k}\frac{g^\mathrm{xc}_{ok}}{\omega_{ok}}\ket{o,o}\bra{o,o}(b^\dag_{ok}-b_{ok})
     \Bigg),
     \label{eq:polaron}
\end{multline}
which displaces bath modes and transforms the entire Hamiltonian ($H=H_\mathrm{S}+H_\mathrm{B}+H_\mathrm{SB}$) into the polaron-transformed version, $\tilde{H} = e^S He^{-S} = \tilde{H}_\mathrm{S} + \tilde{H}_\mathrm{B} + \tilde{H}_\mathrm{SB}$, given in sec.~S1 of the SI.

\subsection{Polaron states}

The polaron states are the eigenstates of the polaron-transformed system Hamiltonian $\tilde{H}_\mathrm{S}$ and represent the simultaneous state of the delocalised electron and hole in the polaron frame (\cref{fig:model}b). 

The polaron transformation leads to spatially smaller states, which are computationally more manageable. This occurs because the transformation decreases the electronic couplings, reducing eigenstate delocalisation~\cite{Rice2018}. 
We quantify the delocalisation of a polaron state $\ket{\nu}$ using the inverse participation ratio (IPR)
\begin{equation}
     \label{eq:IPR}
     \mathrm{IPR}_\nu = \Bigg(\sum_{m,n} \abs{\braket{m,n|\nu}}^4\Bigg)^{-1}, 
\end{equation}
a measure of how many site-pairs $\ket{m,n}$ the polaron state is spread across. The extent of delocalisation is determined by the parameters in the model: electronic couplings and transition dipole moments increase IPRs, while disorder and system-bath coupling decrease them. 

We classify the polaron states into categories based on their overlaps with the different types of site-pairs shown in \cref{fig:cartoon_mechanism}b: exciton site-pairs, bound site-pairs (which are either CT site-pairs or weakly bound site-pairs), and separated site-pairs. Every polaron has an overlap with each of these types of site-pairs; for example, the exciton character of polaron $\ket{\nu}$ is $\sum_{o}\abs{\braket{o,o|\nu}}^2$ and its separated-charge character is $\sum_{m,n}\abs{\braket{m,n|\nu}}^2$, where the sum goes over site-pairs $\ket{m,n}$ in which the electron-hole separation exceeds $r_\mathrm{sep}=\SI{5}{nm}$.
We classify the polaron states by which type of site-pair they overlap with the most. For example, exciton states have most of their population on exciton site-pairs. In addition, a state is hybridised if it has more than $25\%$ of its population on both exciton and non-exciton site-pairs.

\begin{figure}
    \centering
    \includegraphics[width=\columnwidth]{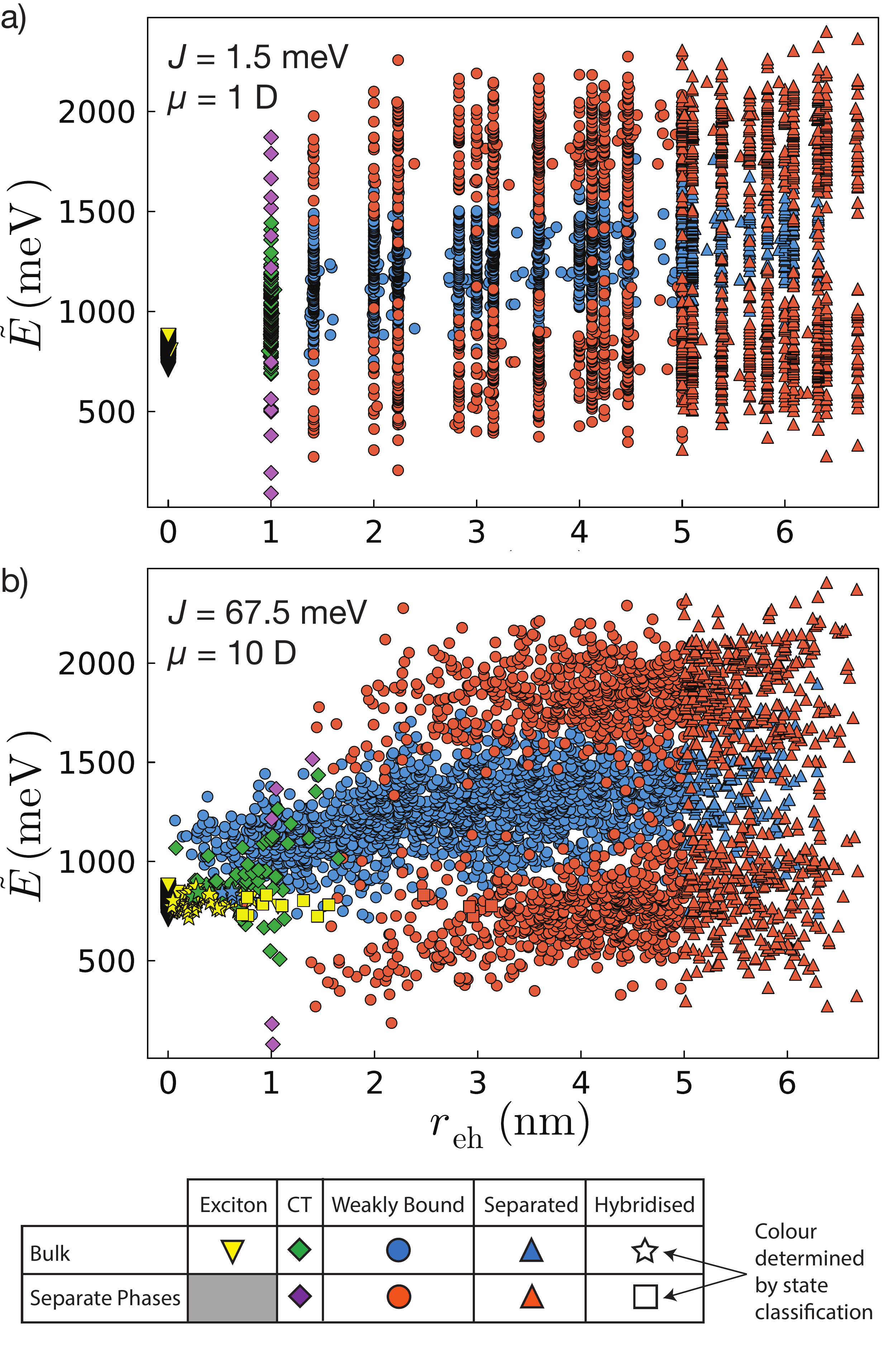}
    \caption{\textbf{Delocalised polaron states.} Energy $\tilde{E}$ and electron-hole separation $r_\mathrm{eh}$ of polaron states of a representative $\tilde{H}_\mathrm{S}$ for 
    \textbf{a)} small couplings and dipole moments ($J=\SI{1.5}{meV}$, $\mu=\SI{1}{D}$) and \textbf{b)} large ones  ($J=\SI{67.5}{meV}$, $\mu=\SI{10}{D}$). States are classified based on which type of site-pair they overlap with the most and whether the electron and the hole in those site-pairs are predominantly in the bulk (same phase) or in separate phases. The two bands of separate-phase states correspond to whether the electron is in the donor and the hole in the acceptor or vice versa.
    For small couplings, the states are localised to about one site-pair centred on the lattice, with no hybridised states forming. When the couplings are larger, states become delocalised, are no longer centred on the lattice, and can be hybridised, with both exciton and non-exciton character. All results are in 2D for $E^\mathrm{LUMO}_\mathrm{offset}=E^\mathrm{HOMO}_\mathrm{offset}=\SI{500}{meV}$, and the other simulation parameters are in sec.~S6 of the SI.
    }
    \label{fig:polaron_states}
\end{figure}

\Cref{fig:polaron_states} shows the polaron states found by diagonalising a representative $\tilde{H}_\mathrm{S}$. At small electron-hole separations, the Coulomb attraction is largest, reducing the energy of the states. 
For small couplings (\cref{fig:polaron_states}a), each state is approximately a site-pair, meaning they are localised close to the lattice. However, for higher couplings (\cref{fig:polaron_states}b), the states become delocalised and are no longer centred close to the lattice. Furthermore, a significant number of hybridised states only forms in the higher-couplings case. The IPRs of the states as a function of energy are included in sec.~S7 of the SI.
 
\subsection{dKMC dynamics}
dKMC propagates the dynamics through hops among the partially delocalised polaron states (\cref{fig:model}c). The polaron transformation allows a perturbative treatment of the dynamics by moving most of the system-bath interaction into the polaron itself. As a result, the polaron interacts relatively weakly with its environment and this remaining coupling can often be treated perturbatively~\cite{Grover1971}. Treating the perturbation to second order leads to the secular polaron-transformed Redfield master equation (sPTRE)~\cite{Lee2015} for transition rates between polaron states, as described in sec.~S2 of the SI.
dKMC is a kinetic Monte Carlo procedure that tracks (and subsequently averages over) individual trajectories through the polaron states using sPTRE rates, in a similar way to how standard KMC uses Marcus or Miller-Abrahams rates to track classical trajectories across sites. The details are given in secs.~S3--S5 of the SI.

\paragraph{Initialisation} We begin a dKMC simulation by generating a disordered energetic landscape and randomly oriented dipoles on each site. Next, we choose the initial exciton state. This involves selecting an excitation location by randomly choosing a point whose distance from the interface is between 1 site and $d_\mathrm{xc}^\mathrm{max}$ (here $\SI{5}{nm}$, motivated by typical domain sizes of $\SI{10}{nm}$). Second, we diagonalise a subset of $\tilde{H}_\mathrm{S}$ that describes a neighbourhood of the selected excitation location that is large enough to contain the states the initial exciton is likely to hop to (\cref{fig:model}c). 
The size of the neighbourhood is calibrated to ensure accuracy as described previously~\cite{Balzer2023} and in sec.~S3 of the SI.
Finally, the initial exciton state is chosen at random among exciton states centred within $\SI{1}{nm}$ of the excitation location, and in proportion to their oscillator strengths, which describe the probability that they would be excited by light. 

\begin{figure*}
    \centering
    \includegraphics[width=\textwidth]{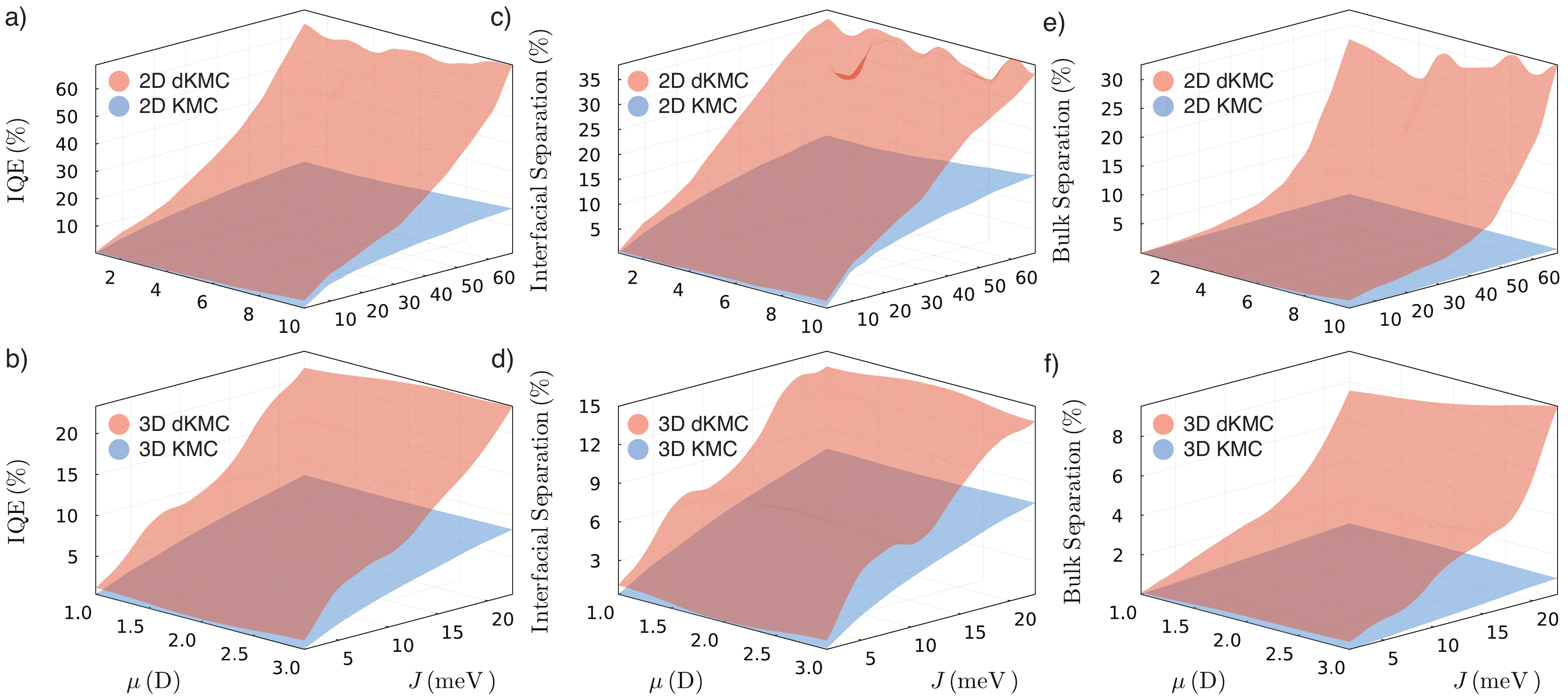}
    \caption{{\bf{Delocalisation enhances charge generation.}}  \textbf{a-b)} IQE of charge generation predicted by KMC~(blue) and dKMC~(orange) as a function of electronic coupling $J$ and transition dipole moment $\mu$ in (a) 2D and (b) 3D. As $J$ and $\mu$ increase, and the charges and excitons become more delocalised, the IQE predicted by dKMC increases compared to KMC, showing that both charge and exciton delocalisation improve charge generation, increasing the IQE by up to $5\times$ in 2D, and $11\times$ in 3D. For both 2D and 3D results, the IQE is split into additive contributions due to \textbf{c-d)} trajectories where charges separate across the interface (pathways~A and~C) and \textbf{e-f)}~trajectories where they separate entirely in one phase (pathway~B). In both 2D and 3D, delocalisation both helps charges separate across the interface and enables them to separate in the bulk, without the assistance of interfacial energetic offsets. We set $E^\mathrm{LUMO}_\mathrm{offset}=E^\mathrm{HOMO}_\mathrm{offset}=\SI{500}{meV}$ and the other simulation parameters are in sec.~S6 of the SI.
    }
    \label{fig:heterojunction}
\end{figure*}

\paragraph{Propagation} After initialisation, the dynamics is propagated through the polaron states using the dKMC procedure, as detailed in sec.~S5 of the SI. At each step, we calculate the outgoing hopping rates from the current polaron state. To do so, we only consider hops to states within a certain distance, allowing for different distances for charge and exciton hops. These cutoffs are calibrated as described previously~\cite{Balzer2023} and in sec.~S3 of the SI.

At each step, the polaron state can also recombine due to its overlap with exciton site-pairs or CT site-pairs. The expressions for these rates are given in sec.~S4 of the SI, and take the form of constant recombination rates modified by the square of amplitudes on site-pairs that can recombine, as in previous work~\cite{Tempelaar2016,Taylor2018,Balzer2022}.  

As in KMC, one hopping destination (including recombination) is chosen probabalistically in proportion to the size of the corresponding rates. The current state and the elapsed time are updated, before diagonalising a new subset of $\tilde{H}_\mathrm{S}$, centred on the new state. 

\paragraph{Termination} The delocalised hopping is repeated until the charges either separate or recombine. The condition for separation is that the expectation value of the electron-hole separation, $\abs{\braket{\nu|\mathbf{r}_\mathrm{e}-\mathbf{r}_\mathrm{h}|\nu}}$, exceeds $r_\mathrm{sep}=\SI{5}{nm}$. For numerical purposes, we also impose a cutoff in the maximum number of hops, $n_\mathrm{hops}$, to prevent infinite loops between, for example, two low-lying traps. This cutoff is chosen to be sufficiently large so that no more than 5\% of trajectories exceed the cutoff, a tolerance that could be reduced by increasing $n_\mathrm{hops}$.

The complete algorithm above is repeated for $n_\mathrm{traj}$ trajectories and on $n_\mathrm{iters}$ realisations of disorder. These numbers determine the uncertainty of the calculations and, in our calculations, are chosen so that the standard error of the mean is never more than 2\% (the values are listed in sec.~S6 of the SI) . Finally, the IQE is calculated as the proportion of trajectories that separate.

\subsection{Pathways of delocalised charge generation}

As a framework for analysing our results, we describe the three pathways for charge generation, illustrated in \cref{fig:cartoon_mechanism}.
Charge generation is the full conversion of excitons into separated charges, and can be divided into two steps (\cref{fig:cartoon_mechanism}b). Exciton dissociation, involves an exciton dissociating into a bound state, either a CT state or a weakly bound state. After that, charge separation is the separation of a bound state into separated charges.

Both steps---exciton dissociation and charge separation---can occur using the interface or entirely in the bulk of a single phase. The combinations of these possibilities lead to the three charge-generation pathways in \cref{fig:cartoon_mechanism}a. Interfacial charge generation (pathway A) is the conventional pathway involving interfacial exciton dissociation and interfacial charge separation, i.e., the exciton dissociates at the interface and the charges then separate in the different phases. In bulk charge generation (pathway B), both the exciton dissociation and the charge separation occur in the same phase. Finally, mixed charge generation (pathway C) involves bulk exciton dissociation followed by interfacial charge separation. A combination involving interfacial exciton dissociation and bulk charge separation is not possible because interfacial exciton dissociation yields charges on different sides of the interface, whose separation cannot occur in the bulk of one phase.

\section{Results and discussion}

\subsection{Delocalisation increases the IQE}

To determine the effect of delocalisation on charge generation, we compare the IQE predicted by standard KMC (which ignores delocalisation) to that predicted by dKMC (which includes it). \Cref{fig:heterojunction}a--b shows the two IQEs as a function of the electronic coupling $J$ and the transition dipole moment~$\mu$; these parameters independently increase the delocalisation of charges and excitons (other parameters being fixed), allowing the effect of charge and exciton delocalisation to be independently assessed. At small $J$ and $\mu$, the IQEs predicted by dKMC and KMC agree, as expected in the limit of localised states. As $J$ and $\mu$ increase, the IQE predicted by dKMC increases compared to classical KMC, demonstrating that both charge and exciton delocalisation improve charge generation. Over our parameter ranges, we see up to a $5\times$ enhancement in 2D, and as much as $11\times$ in 3D, despite the more modest $J$ and $\mu$.

\subsection{Delocalisation enables charge generation without energetic offsets}

Delocalisation enhances the IQE both by enhancing charge generation involving the interface and by enabling charge generation in the bulk of a single phase. In \cref{fig:heterojunction}, we split the IQE into contributions due to interfacial charge separation---trajectories that separate across the interface, so that the electron and hole are on opposite sides (pathways~A and~C, \cref{fig:heterojunction}c--d)---from those due to bulk charge separation---trajectories that separate completely in one phase, with the electron and hole on the same side of the interface (pathway~B, \cref{fig:heterojunction}e--f). As $J$ and $\mu$ increase, and the charges and excitons become more delocalised, dKMC predicts an enhancement in the number of trajectories separating across the interface compared to KMC (up to $3\times$ in 2D and $2\times$ in 3D for our parameter choices), demonstrating that both charge and exciton delocalisation are beneficial at improving interfacial charge separation. However, \cref{fig:heterojunction}e--f reveals that delocalisation also enables charges to separate in the bulk of one phase via pathway~B, i.e., without the assistance of an interfacial energetic offset. In both 2D and 3D, KMC predicts a negligible bulk-separation IQE, as expected for localised charges, which struggle to overcome the exciton binding energy without the interfacial energetic offset providing a driving force. However, as $J$ and $\mu$ increase and the states become delocalised, dKMC shows that excitons are able to separate efficiently within a single material, especially in 3D. Compared to KMC, there is an enhancement of up to $47\times$ in 2D and already $12\times$ in 3D for the smaller $J$ and $\mu$. Therefore a large proportion of delocalised charge generation occurs in the bulk without the assistance of an interfacial offset.

\begin{figure}
    \centering
    \includegraphics[width=0.9\columnwidth]{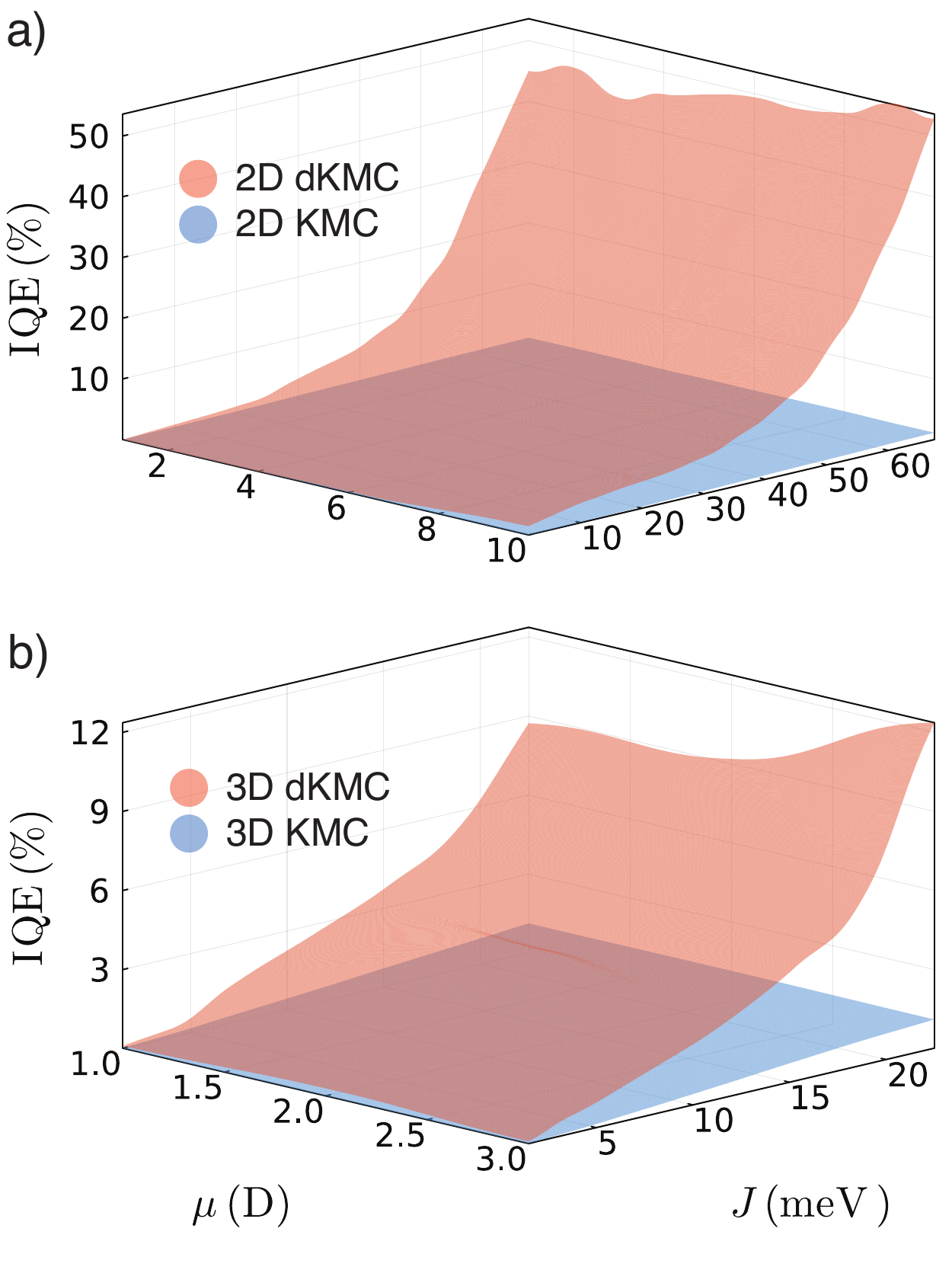}
    \vspace{-3mm}
    \caption{{\bf{Delocalisation enables charge generation without energetic offsets.}} IQE of charge generation in a homojunction (no interfacial energetic offset) predicted by KMC (blue) and dKMC (orange) as a function of the electronic coupling $J$ and transition dipole moment $\mu$ in \textbf{(a)}~2D \textbf{(b)}~and 3D. Both charge and exciton delocalisation improve the efficiency, especially in 3D. Simulation parameters are in sec.~S6 of the SI, except that $E^\mathrm{LUMO}_\mathrm{offset}=E^\mathrm{HOMO}_\mathrm{offset}=0$.
    }
    \label{fig:homojunction}
\end{figure}

Indeed, delocalisation enables efficient charge generation even in a single phase (a homojunction), without energetic offset to assist in separation. \Cref{fig:homojunction} shows the IQE for a homojunction, with KMC again unable to predict efficient separation at any $J$ and $\mu$. However, when the states are delocalised, at high $J$ and $\mu$, dKMC predicts considerable IQEs, with enhancements of $48\times$ in 2D, and already $11\times$ in 3D for modest $J$ and $\mu$. This demonstration that delocalisation can enable charge generation in a homojunction could explain observations of charge generation in neat Y6~\cite{Wang2020,Price2022,Saglamkaya2023}.

\clearpage

\subsection{Energetic offsets are still beneficial for delocalised charge generation}

\begin{figure}[b]
    \centering
    \includegraphics[width=0.9\columnwidth]{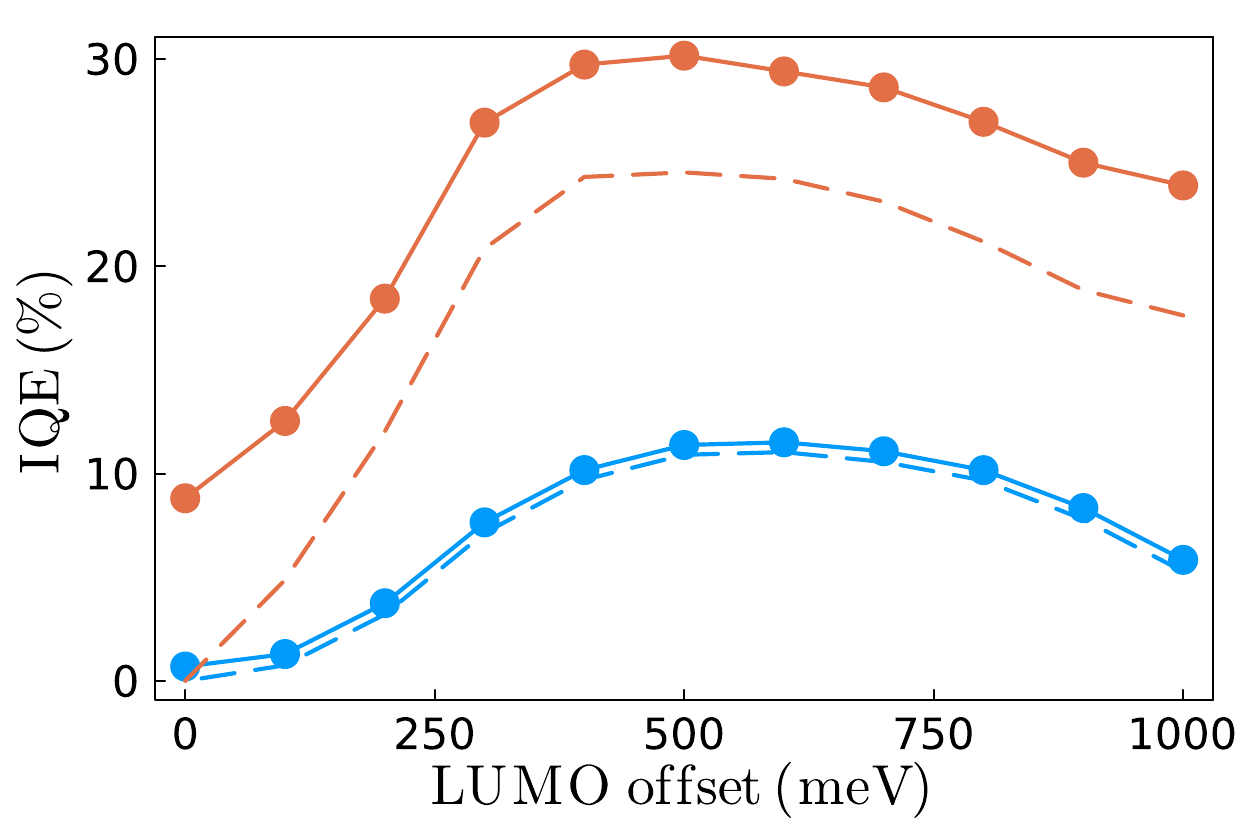}
    \caption{{\textbf{Interfacial energetic offsets remain beneficial with delocalisation.}} IQE predicted by KMC (blue dots) and dKMC (orange dots) in 2D. Dashed lines are the contributions to the IQE from trajectories where charges separate across the interface (pathways~A and~C), meaning that the difference between the solid and dashed lines is due to trajectories that separate in the bulk (pathway~B). With KMC, large energetic offsets are required to see significant charge generation. By contrast, with dKMC, delocalisation enables charge generation in the bulk without interfacial energetic offsets. However, even though the interface is not required, it still improves interfacial charge generation by providing a driving force to dissociate excitons and by further separating states that have dissociated in the bulk but not fully separated. We set $J=\SI{45}{meV}$ and $\mu=\SI{5}{D}$, and the other simulation parameters are in sec.~S6 of the SI.}
    \label{fig:LUMO_offsets}
\end{figure}

Of course, interfacial energetic offsets are beneficial even with delocalisation because they can still drive interfacial charge separation. \Cref{fig:LUMO_offsets} shows heterojunction IQEs for varying LUMO offset, finding that for KMC to predict significant charge generation, a large offset is required. By contrast, dKMC predicts charge generation at all offsets, including small ones and even zero (as in the homojunctions above), a finding that could help explain efficient low-offset NFA devices~\cite{Bin2016,Hou2018,Perdigon2020}.

\begin{figure*}
    \centering
    \includegraphics[width=\textwidth]{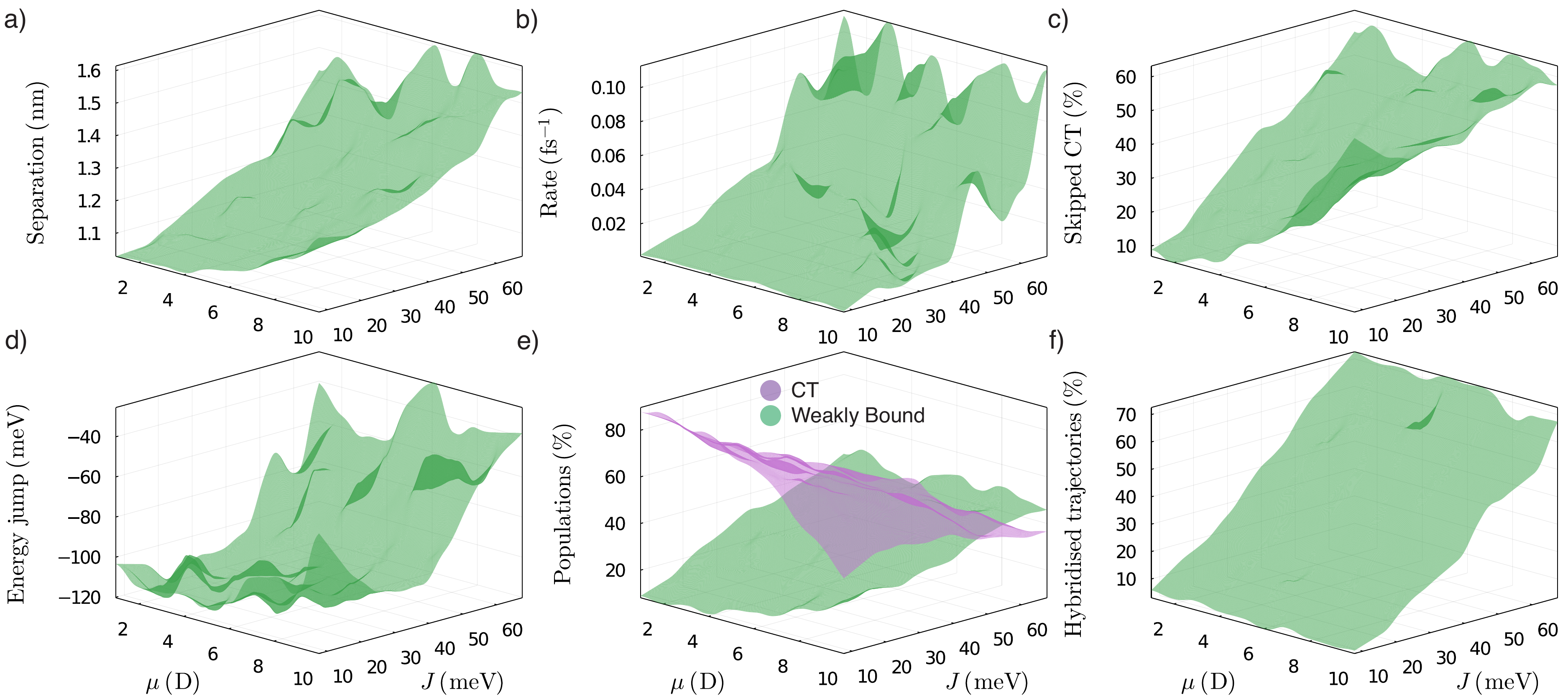}
    \caption{\textbf{The mechanism of delocalisation enhancement of exciton dissociation.} Delocalisation enables charges to hop further and faster out of an exciton, allowing them to skip CT states and use hybridised states.
    \textbf{a)}~As delocalisation increases with electronic coupling $J$ and transition dipole moment $\mu$, charges hop further out of the exciton. Shown is the mean electron-hole separation in the first state following exciton dissociation.
    \textbf{b)} Excitons dissociate faster with delocalised states, as shown by the rate of the exciton dissociation event.
    \textbf{c)}~Because of longer-range hopping, the fraction of separated trajectories that skip CT states increases with the couplings. 
    \textbf{d)}~By skipping CT states, trajectories that are more delocalised can avoid large downhill hops, meaning they are less likely to become trapped in low-energy CT states. Shown is the average energy lost during the exciton dissociation event.
    \textbf{e)}~As couplings increase, and states become more delocalised, the character of the first bound state following exciton dissociation changes from CT state to weakly bound state. Shown are the mean populations of this state on CT and weakly bound site-pairs. 
    \textbf{f)}~With stronger couplings, exciton dissociation is more likely to involve a hybridised state. Shown is the fraction of separated trajectories where exciton dissociation involves a state with more than 25\% of both exciton and non-exciton character.
    All results are in 2D for $E^\mathrm{LUMO}_\mathrm{offset}=E^\mathrm{HOMO}_\mathrm{offset}=\SI{500}{meV}$, and the other simulation parameters are in sec.~S6 of the SI.
    }
    \label{fig:mechanism}
\end{figure*}

However, \cref{fig:LUMO_offsets} also shows that driving forces are still helpful for delocalised charge generation by improving separation across the interface. This improvement arises for two reasons.

First, as in conventional theories, an energetic offset helps dissociate excitons that would otherwise recombine in the bulk. Our results share another feature of classical theories based on Marcus theory, the inverted regime at very high offsets, where the transition is so far energetically downhill that the charge-transfer rates across the interface decrease.

The second benefit of the offset, which is not seen in conventional theories, is enabling mixed charge generation (pathway C in \cref{fig:cartoon_mechanism}). Mixed charge generation occurs when an exciton dissociates in the bulk to a bound state without fully separating, and the bound state then uses an interface for the charges to separate completely. Our finding that delocalisation enables mixed charge generation supports the hypothesis of delocalised, partially separated intermediate states forming in single domains before they reach the interface~\cite{Wang2020,Tu2020,Zhu2021,Zhang2020,Dimitriev2022,Li2023,Li2023_2}.

\subsection{Mechanism of delocalisation enhancements}

Microscopically, the enhancements above arise because delocalisation enables longer and faster hops, which can bypass CT states and involve hybridised states instead. To see this, we note that delocalisation can improve overall charge generation by enhancing either the exciton-dissociation or the charge-separation steps in \cref{fig:cartoon_mechanism}b. Previously, we described that delocalisation benefits charge separation by increasing the overlaps between states to kinetically assist charges in overcoming their Coulomb attraction~\cite{Balzer2022}. And once separated, delocalised charges are more easily extracted because the longer and faster hops increase their mobilities~\cite{Balzer2020}. Having extended dKMC to include excitons, we can now show that delocalisation benefits exciton dissociation in the same way, by accelerating long-distance hopping.

To understand the delocalisation enhancement of exciton dissociation, we study the exciton-dissociation event, the last hop in each trajectory out of an exciton state and into a non-exciton one. To do so, we identify trajectories where charges have separated, find the hop where the state character changed from predominantly exciton to non-exciton, and then analyse the two states immediately before and after the hop. In the following analysis, we study both types of exciton-dissociation events---at the interface and in the bulk---together, as the mechanism by which delocalisation improves them is the same (see sec.~S8 of the SI).

Delocalised charges hop further out of an exciton during dissociation (\cref{fig:mechanism}a) and they do so faster (\cref{fig:mechanism}b). \Cref{fig:mechanism}a shows that when charges are localised, or when classical KMC is used, the separation after leaving an exciton state is 1 lattice spacing, as expected for charges forming a tightly bound CT state. For dKMC, as the states become more delocalised with increasing couplings, the charges can hop further out of the exciton state, causing $r_\mathrm{eh}$ to exceed 1 lattice spacing. The ability to hop further provides the additional benefit of more possible hopping destinations, which can be important in disordered materials and when dissociating an exciton in systems with little or no energetic offset. The larger overlaps between delocalised states also increase the speed of individual hops, allowing for faster hopping out of the exciton state (\cref{fig:mechanism}b).

The longer hopping distance means that charges departing exciton states can often skip CT states, as shown by the proportion of separating trajectories that skip CT states in \cref{fig:mechanism}c. Because CT states tend to be tightly bound, skipping them means that a trajectory can avoid deeply downhill hops, which risk trapping it in a low-energy CT state. As shown in \cref{fig:mechanism}d, the average energy lost upon hopping out of an exciton is indeed less for more delocalised states.

The enhancements above are facilitated by strong couplings, which also change the character of the states involved. As the delocalised charges hop further, the first separated state has less CT character and more weakly bound character, as shown in \cref{fig:mechanism}e. Indeed, the strong couplings cause the formation of hybridised states, whose prevalence increases for more delocalised states (\cref{fig:mechanism}f). This result supports the hypothesis that hybridisation is involved in delocalised charge separation~\cite{Qian2018,Eisner2019,Coropceanu2019,Qian2023}.

\subsection{Outlook}

Our results are the first three-dimensional simulations of charge generation that include disorder, delocalisation, and polaron formation. They are possible because dKMC reduces the computational cost of sPTRE by as much as 54 orders of magnitude (see sec.~S3 of the SI).

However, three-dimensional dKMC remains limited by its computational complexity to moderate couplings and transition dipole moments (see \cref{fig:heterojunction,fig:homojunction}). Future work may extend these calculations using further simplifications to dKMC, such as jKMC~\cite{jkmc}.

Furthermore, dKMC is limited by approximations in the underlying sPTRE, which could be relaxed if necessary, albeit at increased computational expense. For example, sPTRE is accurate in the regimes studied here~\cite{Lee2012,Lee2015}, but can become inaccurate for systems weakly coupled to slow baths~\cite{Lee2012}. To treat such systems, the full polaron transformation could be replaced with a variational version~\cite{Pollock2013,Jang2022}. The secular approximation ignores coherences, which is ordinarily valid because sunlight does not induce coherences~\cite{Jiang1991,Mancal2010,Brumer2012,Kassal2013,Brumer2018,Tomasi2019,Tomasi2020,Tomasi2021} and even if they were induced otherwise, they would not survive on charge-transport timescales. However, if required, coherences could be tracked at additional computational expense by not making the secular approximation and tracking the entire density matrix. Finally, sPTRE uses a local system-bath coupling, as is relevant for disordered molecular systems~\cite{MayKuhn}, but to treat organic crystals, non-local system-bath couplings may need to be included to describe the fluctuations of electronic couplings~\cite{Troisi2006,Troisi2006_Review,Arago2016}.

We expect that using dKMC in multiscale simulations could enable a full device model that incorporates delocalisation. This would involve parameterising dKMC using atomistic simulations, such as molecular mechanics followed by electronic-structure calculations. The results of dKMC could then parameterise drift-diffusion models for computing device properties. We also expect that dKMC can be extended to model other fundamental processes in organic devices, such as singlet fission, which is theoretically similar to charge generation.

\section{Conclusion}
Extending dKMC to the full charge-separation process, from excitons to separated charges, explains observations of efficient charge separation in disordered OPVs, especially those with little to no energetic offset. dKMC reveals that delocalisation produces large IQEs compared to those predicted by classical KMC. Delocalisation does so by enabling separated charges to be generated in the bulk and by enabling charges to hop faster and further out of an exciton during dissociation, mediated by hybridised states. Together with past dKMC studies, our results show that delocalisation accelerates all the fundamental processes in organic electronics---charge transport~\cite{Balzer2020}, exciton diffusion~\cite{Balzer2023}, charge separation~\cite{Balzer2022}, and exciton dissociation (this work)---in essentially the same way, by allowing polaron states to hop further and faster.
We anticipate that this new mechanistic understanding of charge generation in OPVs will explain the performance of efficient low-offset heterojunctions and neat devices and help design more efficient future ones.

\begin{acknowledgments}
We were supported by a Westpac Scholars Trust Future Leaders Scholarship, the Australian Research Council (DP220103584), the Australian Government Research Training Program, and by computational resources from the National Computational Infrastructure (Gadi) and the University of Sydney Informatics Hub (Artemis).
\end{acknowledgments}

\bibliography{bib}

\end{document}


\title{Supporting Information: Delocalisation enables efficient charge generation in organic photovoltaics, even with little to no energetic offset}

\author{Daniel Balzer}
\author{Ivan Kassal*}
\affiliation{School of Chemistry, University of Sydney, NSW 2006, Australia\\
*\;Email: ivan.kassal@sydney.edu.au}

\maketitle

\setcounter{section}{0}
\renewcommand{\thesection}{S\arabic{section}}%
\setcounter{equation}{0}
\renewcommand{\theequation}{S\arabic{equation}}%
\setcounter{figure}{0}
\renewcommand{\thefigure}{S\arabic{figure}}%
\setcounter{table}{0}
\renewcommand{\thetable}{S\arabic{table}}%
\setcounter{algorithm}{0}
\renewcommand{\thealgorithm}{S\arabic{algorithm}}%

\section{Polaron-transformed Hamiltonian}
Applying the polaron transformation (eq. (6) of the main text) to the total Hamiltonian results in the polaron-transformed Hamiltonian
\begin{equation}
    \tilde{H} = e^S He^{-S} = \tilde{H}_\mathrm{S} + \tilde{H}_\mathrm{B} + \tilde{H}_\mathrm{SB}.
\end{equation} 
To evaluate this, we follow previous work~\cite{Lee2015}, but adjust the equations to account for the two-particle picture. The polaron-transformed system Hamiltonian is
\begin{equation}
    \label{eq:polaron_H_S}
    \tilde{H}_\mathrm{S} = \sum_{m,n} \tilde{E}_{mn} \ket{m,n}\bra{m,n}  + \sum_{m \neq m',n} \kappa^\mathrm{e}_{mm'}J_{e} \ket{m,n}\bra{m',n} + \sum_{m,n \neq n'} \kappa^\mathrm{h}_{nn'}J_{h} \ket{m,n}\bra{m,n'}  + \sum_{o \neq o'} \kappa^\mathrm{xc}_{oo'}J_\mathrm{xc} \ket{o,o}\bra{o',o'},
\end{equation}
where the shifted energies are 
\begin{align}
    \tilde{E}_{mn}&=E_{mn}-\sum_k\abs{g^\mathrm{e}_{mk}}^2/\omega_{mk}-\sum_k\abs{g^\mathrm{h}_{nk}}^2/\omega_{nk} \quad\text{when } m\neq n\\
    \tilde{E}_{oo}&=E_{oo}-\sum_k\abs{g^\mathrm{xc}_{ok}}^2/\omega_{ok}
\end{align}
and the couplings are renormalised by factors of the form
\begin{equation}
    \kappa^\mathrm{e/h/xc}_{mm'} = \exp\left(-\frac{1}{2}\sum_k\left(\frac{\left(g^\mathrm{e/h/xc}_{mk}\right)^2}{\omega^2_{mk}}\coth{\frac{\beta\omega_{mk}}{2}}+\frac{\left(g^\mathrm{e/h/xc}_{m'k}\right)^2}{\omega^2_{m'k}}\coth{\frac{\beta\omega_{m'k}}{2}}\right)\right),
\end{equation}
where the superscript $\mathrm{e/h/xc}$ indicates which system-bath coupling strength $g^\mathrm{e/h/xc}_{mk}$ is used and $\beta=1/k_\mathrm{B} T$ (here, we assume $T=\SI{300}{K}$).
The bath Hamiltonian is unchanged, $\tilde{H}_\mathrm{B}=H_\mathrm{B}$, while the polaron-transformed interaction Hamiltonian is
\begin{equation}
    \label{eq:polaron_H_SB}
    \tilde{H}_\mathrm {SB} = \sum_{m \neq m',n} J_{e} \ket{m,n}\bra{m',n}V^\mathrm{e}_{mm'}  + \sum_{m,n \neq n'}J_{h} \ket{m,n}\bra{m,n'}V^\mathrm{h}_{nn'} + \sum_{m,n \neq n'}J_\mathrm{xc} \ket{o,o}\bra{o,o'}V^\mathrm{xc}_{oo'},
\end{equation}
where
\begin{equation}
    V^\mathrm{e/h/xc}_{mm'} = \exp{\left(\sum_k\frac{g^\mathrm{e/h/xc}_{mk}}{\omega_{mk}}\left(b^\dag_{mk}-b_{mk}\right)\right)} \exp{\left(-\sum_k\frac{g^\mathrm{e/h/xc}_{m'k}}{\omega_{m'k}}\left(b^\dag_{m'k}-b_{m'k}\right)\right)}- \kappa^\mathrm{e/h/xc}_{mm'}.
\end{equation}

To simplify the system-bath interaction, which currently involves a sum over bath modes, we assume that all sites couples to their own baths with identical strengths ($g^\mathrm{e}_{nk}=g^\mathrm{e}_k$, $g^\mathrm{h}_{nk}=g^\mathrm{h}_k$, and $g^\mathrm{xc}_{nk}=g^\mathrm{xc}_k$), and that the discrete spectral density 
\begin{equation}
J^\mathrm{e/h/xc}(\omega)=\sum_k \left(g^\mathrm{e/h/xc}_k\right)^2\delta(\omega-\omega_k)
\end{equation}
can be replaced with a continuous one. We apply the commonly used super-Ohmic spectral density~\cite{Jang2002,Jang2011,Pollock2013,Lee2015}
\begin{equation}
    J^\mathrm{e/h/xc}(\omega) = \frac{\lambda^\mathrm{e/h/xc}}{2} \left(\frac{\omega}{\omega^\mathrm{e/h/xc}_c}\right)^3 \exp\left(-\frac{\omega}{\omega^\mathrm{e/h/xc}_c}\right)\\
\end{equation}
but it would be possible to choose more structured versions for specific organic semiconductors. These approximations reduce $\kappa^\mathrm{e/h/xc}_{mm'}$ to 
\begin{equation}
     \kappa^\mathrm{e/h/xc}_{mm'} = \kappa^\mathrm{e/h/xc} = \exp\left(-\int_0^\infty d\omega\frac{J^\mathrm{e/h/xc}(\omega)}{\omega^2}\coth{\frac{\beta\omega}{2}}\right).
\end{equation}

\section{Polaron-transformed Redfield equation}
The polaron transformation reduces the system-bath coupling by moving most of the interaction into the polaron frame. This allows the remaining system-bath interaction ($\tilde{H}_\mathrm{SB}$) to be treated perturbatively to second order, resulting in the secular polaron-transformed Redfield equation (sPTRE)~\cite{Lee2015}. The equations that follow are analogous to those derived for sPTRE~\cite{Lee2015}, but they have been generalised for the two-particle picture. The quantum master equation describes the time evolution of polaron state populations,
\begin{equation}
     \label{eq:sPTRE_master_equation}
     \frac{d\rho_{\nu}(t)}{dt} = \sum_{\nu'}R_{\nu\nu'}\rho_{\nu'}(t),
\end{equation}
where the secular Redfield tensor contains rates of population transfer between every pair of polaron states. The Redfield tensor
\begin{equation}
     \label{eq:secular_redfield_tensor}
     R_{\nu\nu'} = 2\Re\Big(\Gamma_{\nu'\nu,\nu\nu'} - \delta_{\nu\nu'}\sum_\kappa\Gamma_{\nu\kappa,\kappa\nu'}\Big),
\end{equation}
consists of damping rates 
\begin{multline}
    \label{eq:Gamma}
    \Gamma_{\mu\nu,\mu'\nu'} = 
    \sum_{n}\sum_{p,q,p',q'} J^\mathrm{e}_{pn,qn} J^\mathrm{e}_{p'n,q'n} \braket{\mu|p,n}
     \braket{q,n|\nu}\braket{\mu'|p',n}\braket{q',n|\nu'} K^\mathrm{e}_{pq,p'q'}(\omega_{\nu'\mu'}) \\
     + \sum_{m}\sum_{p,q,p',q'} J^\mathrm{h}_{mp,mq} J^\mathrm{h}_{mp',mq'} \braket{\mu|m,p}  \braket{m,q|\nu}\braket{\mu'|m,p'}\braket{m,q'|\nu'}K^\mathrm{h}_{pq,p'q'}(\omega_{\nu'\mu'}) \\ 
    +  \sum_{p,q,p',q'} J^\mathrm{xc}_{pp,qq} J^\mathrm{xc}_{p'p',q'q'} \braket{\mu|p,p} \braket{q,q|\nu}\braket{\mu'|p',p'}\braket{q',q'|\nu'} K^\mathrm{xc}_{pq,p'q'}(\omega_{\nu'\mu'}),
 \end{multline}
where the sums go over sites $m,n,p,q,p',q'$, where $\omega_{\nu'\mu'}=E_{\nu'}-E_{\mu'}$, and where
\begin{equation}
    K^\mathrm{e/h/xc}_{pq,p'q'}(\omega) = \int_0^\infty e^{i\omega \tau}\braket{V^\mathrm{e/h/xc}_{pq}(\tau)\tilde{\hat{V}}^\mathrm{e/h/xc}_{p'q'}(0)}_B d\tau.
\end{equation}
This integral contains the bath correlation function~\cite{Jang2011}
\begin{equation}
    \braket{V^\mathrm{e/h/xc}_{pq}(\tau)V^\mathrm{e/h/xc}_{p'q'}(0)}_B = \left(\kappa^\mathrm{e/h/xc}\right)^2\left(e^{\lambda_{pq,p'q'}\phi^\mathrm{e/h/xc}(\tau)} - 1\right),
\end{equation}
where $\lambda_{pq,p'q'} = \delta_{pp'} - \delta_{pq'} + \delta_{qq'} - \delta_{qp'}$, and 
\begin{equation}
    \phi^\mathrm{e/h/xc}(\tau) = \int_0^\infty d\omega\frac{J^\mathrm{e/h/xc}(\omega)}{\omega^2}\big(\cos(\omega \tau)\coth(\beta\omega/2) - i \sin(\omega \tau)\big).
\end{equation}

\newpage

\section{dKMC approximations}

\begin{figure*}[b]
     \centering
     \includegraphics[width=\textwidth]{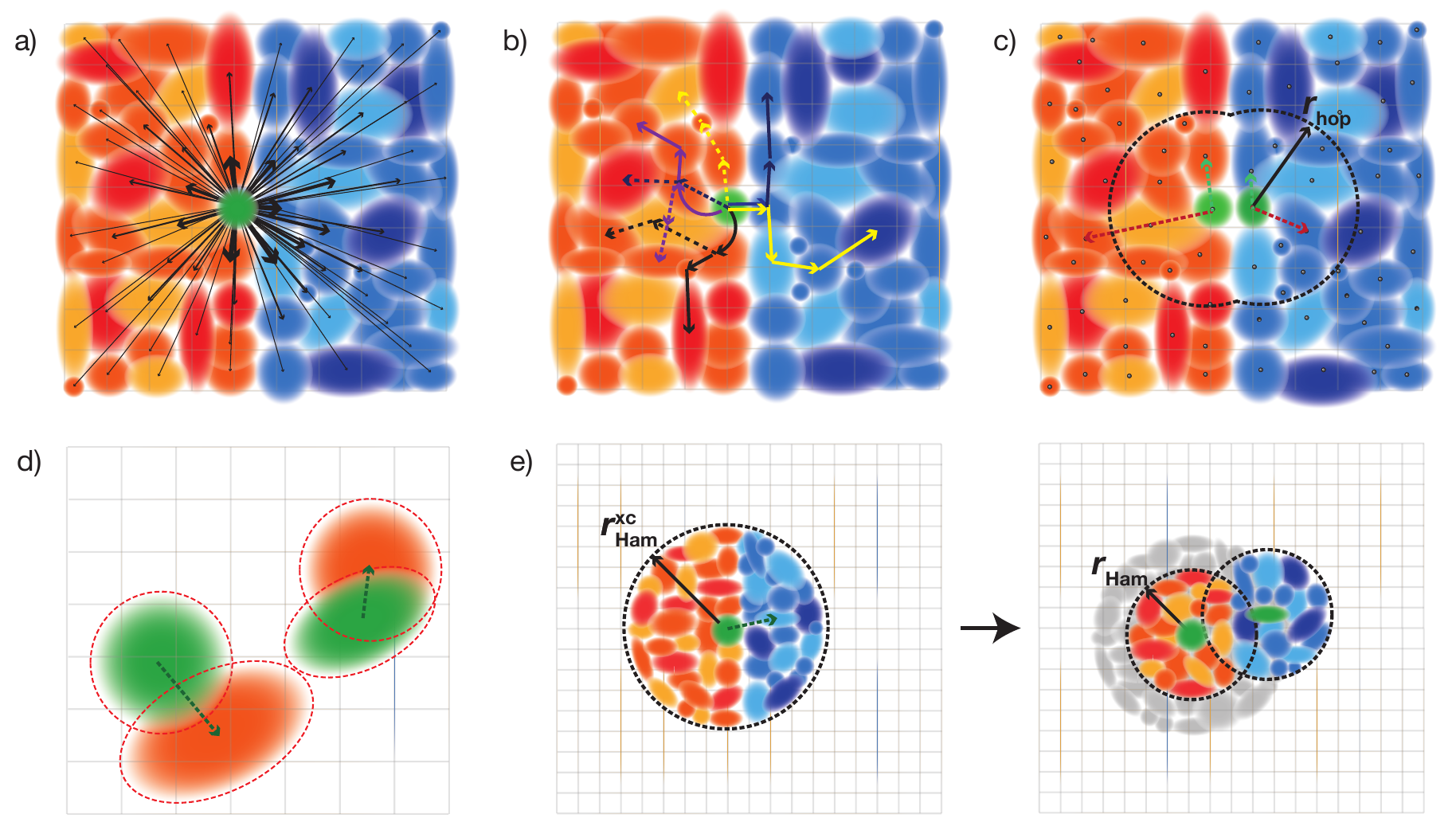}
     \caption{\textbf{Approximations of dKMC for charge generation.} 
     \textbf{a)} The high computational cost of sPTRE arises from the need to track all possible transitions (black arrows) in the time-dependent evolution of the populations of all polaron states. To avoid this cost, dKMC makes the following four approximations.
     \textbf{b)} Kinetic Monte Carlo: we map the quantum master equation onto KMC, which propagates and averages many individual trajectories formed from sequential hops from the current state (in green), chosen probabilistically. Straight arrows denote electron (solid) and hole (dashed) hops, while curved arrows denote exciton hops.
     \textbf{c)} Hopping radius: we only calculate hopping rates to polaron states where the electron and the hole, or the exciton, are sufficiently close to their current positions. In the diagram, the hop depicted by the green arrows is allowed, and the red one is not.
     \textbf{d)} Population cutoff: in each rate calculation from the current state (both the electron and the hole depicted in green) to a possible destination state (in orange), we ignore contributions from site-pairs whose overlap with the initial or final states is insignificant, depicted as lattice points outside the dotted red outlines.
     \textbf{e)} Diagonalising on the fly: rather than calculating all polaron states, we only calculate them for small subsystems surrounding the electron and hole, or exciton, after each hop.}
     \label{fig:approximations}
 \end{figure*}

The sPTRE master equation is too computationally expensive to apply to the full two-particle charge separation problem, so we apply four approximations to transform the method to dKMC. These follow our previous derivations of dKMC~\cite{Balzer2020,Balzer2022,Balzer2023}, but are adapted to the two-body situation. Tracking the full time-evolution according to the sPTRE (\cref{fig:approximations}a) has three computationally difficult steps that dKMC overcomes. First, calculating all $N^{2d}$ polaron states involves diagonalising the $N^{2d}\times N^{2d}$ polaron-transformed system Hamiltonian $\tilde{H}_\mathrm{S}$, a task that scales as $O(N^{6d})$. Second, the full polaron-transformed Redfield tensor involves calculating $N^{4d}$ rates between all pairs of $N^{2d}$ polaron states. Finally, calculating each rate involves a sum over $N^{8d}$ quadruples of site-pairs. In total, the sPTRE procedure scales as $O(N^{6d})+O(N^{12d})$.

The first approximation is mapping the quantum master equation onto KMC (\cref{fig:approximations}b). On each of the $n_\mathrm{iters}$ realisations of disorder, rather than tracking the evolution of the full density matrix using sPTRE, we use the KMC procedure to track $n_\mathrm{traj}$ individual stochastic trajectories, which are then averaged. Each trajectory is propagated by choosing the next state probabilistically using the outgoing rates of population transfer from the current state. Therefore, instead of calculating rates between all pairs of states as in sPTRE, KMC only calculates outgoing rates at each of the $n_\mathrm{hops}$ hops, reducing the number of calculated rates from $O(N^{4d})$ to $O(N^{2d}n_\mathrm{hops}n_\mathrm{traj})$.

The second approximation is the use of hopping cutoffs (\cref{fig:approximations}c). Instead of calculating all outgoing rates at each hop, we only calculate rates to states that are close enough to the current state. In previous work, we calibrated a charge hopping radius $r_\mathrm{hop}$~\cite{Balzer2020} and an exciton hopping radius $r_\mathrm{hop}^\mathrm{xc}$~\cite{Balzer2023}, which are the distances required to contain, on average, enough states such that the rates to them comprise more than a fraction $a_\mathrm{dKMC}$ of all rates (we set $a_\mathrm{dKMC}=0.99$). In the two-particle picture, we only calculate rates from state $\nu$ to states $\nu'$ if the hop would displace the electron and the hole by a total of less than $r_\mathrm{hop}$ in expectation value. That is, we only consider those destination states $\nu'$ for which $\abs{\vec{C}^\mathrm{e}_\nu-\vec{C}^\mathrm{e}_{\nu'}} + \abs{\vec{C}^\mathrm{h}_\nu-\vec{C}^\mathrm{h}_{\nu'}} < r_\mathrm{hop}$, where the expectation value of the electron's position is $\vec{C}^\mathrm{e}_\nu=\bra{\nu}\vec{r}^\mathrm{e}\ket{\nu}$ and likewise for the hole.
In addition, if the current state is an exciton state, we also calculate rates to all exciton states whose position $\vec{C}^\mathrm{xc}_{\nu'}=\bra{\nu'}\vec{r}^\mathrm{xc}\ket{\nu'}$ is within $r_\mathrm{hop}^\mathrm{xc}$ of the current state's position $\vec{C}^\mathrm{xc}_{\nu}$. This is required because exciton states are typically more delocalised and can hop further due to longer-range dipole-dipole coupling. Overall, imposing the hopping cutoffs reduces the number of rates calculated at each hop from $O(N^{2d})$ to $O(r_\mathrm{hop}^{2d})$.

The third approximation is including a site-pair contribution cutoff when calculating each hopping rate (\cref{fig:approximations}d). Instead of including all site-pairs in the sums defining the damping rates of \cref{eq:Gamma}, we ignore contributions of site-pairs with insignificant overlaps with either the initial or final states. For each summation index of \cref{eq:Gamma}, we only include the minimum number of site-pairs $(m,n)$ that together support at least a fraction $a_\mathrm{dKMC}$ of the population of $\nu$ or $\nu'$, as appropriate 
(we set $a_\mathrm{dKMC}=0.99$). Ignoring contributions from site-pairs with small amplitude significantly reduces the cost of calculating each rate. However, as the spatial extent of a state is difficult to predict in general, estimating the reduction in scaling is difficult. By assuming that the each state is appreciably delocalised across not more than than $r_\mathrm{hop}^{2d}$ site-pairs, the cost for each rate calculation is no more than $O(r_\mathrm{hop}^{8d})$, down from $O(N^{8d})$.

The fourth and final approximation is diagonalising $\tilde{H}_S$ on the fly (\cref{fig:approximations}e). Instead of diagonalising the Hamiltonian describing all site-pairs, we diagonalise Hamiltonians representing subsystems, i.e., only containing site-pairs in which both the electron and the hole are close enough to their current locations. After every hop, a new subset of the original $\tilde{H}_S$ is diagonalised for the subsystem surrounding the new locations of the charges. In previous work, we calibrated Hamiltonian radii $r_\mathrm{Ham}$ and  $r_\mathrm{Ham}^\mathrm{xc}$, which represent how large a subsystem Hamiltonian is required to calculate outgoing rates to a desired accuracy $a_\mathrm{dKMC}$~\cite{Balzer2020,Balzer2023}. In the two-particle picture, we only include site-pairs $(m,n)$ whose combined distance from the current locations of the electron and the hole is within $r_\mathrm{Ham}$, $\abs{\vec{r}_m-\vec{C}^\mathrm{e}_{\nu}} + \abs{\vec{r}_n-\vec{C}^\mathrm{h}_{\nu}} < r_\mathrm{Ham}$. In addition, if the current state is classified as an exciton, we also include exciton site-pairs $(o,o)$ within a larger exciton Hamiltonian radius $r_\mathrm{Ham}^\mathrm{xc}$ of the exciton, i.e., $\abs{\vec{r}_o-\vec{C}^\mathrm{xc}_{\nu}}< r_\mathrm{Ham}^\mathrm{xc}$. Only diagonalising subsystems reduces the cost of finding the required polaron states from $O(N^{6d})$ to $O(r_\mathrm{Ham}^{6d}n_\mathrm{hops}n_\mathrm{traj})$. In practice, we calculate the Hamiltonian and hopping radii simultaneously, using the same accuracy (here chosen to be $a_\mathrm{dKMC}=0.99$), using the procedure outlined in previous work~\cite{Balzer2023}.

Overall, the four approximations described above reduce the scaling from $O(N^{6d})+O(N^{12d})$ to $O(r_\mathrm{Ham}^{6d}n_\mathrm{hops}n_\mathrm{traj})+O(n_\mathrm{hops}n_\mathrm{traj}r_{hop}^{2d}r_\mathrm{hop}^{8d})$. For exmaple, for the largest system we study ($d=3$, $N=100$, $\mu=\SI{3}{D}$, $J=\SI{22.5}{meV}$), dKMC reduces the computational complexity of calculations by 54 orders of magnitude over sPTRE.

\section{Recombination}
Recombination of excitons and CT states is typically treated using constant rates $R_\mathrm{recomb}^\mathrm{xc}$ and $R_\mathrm{recomb}^\mathrm{CT}$ in classical kinetic Monte Carlo. That is, whenever the electron and the hole are on the same site, they can recombine with rate $R_\mathrm{recomb}^\mathrm{xc}$, and when they are on neighbouring sites, they can recombine with rate $R_\mathrm{recomb}^\mathrm{CT}$. 

In dKMC, we use Fermi's golden rule to calculate the corresponding exciton and CT recombination rates from a delocalised polaron state $\nu$~\cite{Balzer2022}. The golden-rule rate for exciton recombination,
 \begin{equation}
k^\nu_\mathrm{xc,recomb}=2\pi\Bigg|\sum_{o}\braket{\nu|o,o}\braket{o,o|H|g}\Bigg|^2\rho_\mathrm{recomb},
\end{equation}
involves a sum over exciton site-pairs (where the electron and hole are on the same site) and the density of states $\rho_\mathrm{recomb}$. If the exciton site-pairs are all equally coupled to the ground state with strength $\braket{m,n|H|g}=J_\mathrm{recomb}^\mathrm{xc}$, the exciton recombination rate becomes the standard Monte-Carlo rate adjusted by a delocalisation correction,
 \begin{equation}
     \label{eq:xc_recomb}
     k^\nu_\mathrm{xc,recomb}= R_\mathrm{recomb}^\mathrm{xc}\Bigg|\sum_{o}\braket{\nu|o,o}\Bigg|^2,
\end{equation}
where $R_\mathrm{recomb}^\mathrm{xc}=2\pi|J_\mathrm{recomb}^\mathrm{xc}|^2\rho_\mathrm{recomb}$. 

The same approach for CT-state recombination leads to  
 \begin{equation}
     \label{eq:CT_recomb}
     k^\nu_\mathrm{CT,recomb}= R_\mathrm{recomb}^\mathrm{CT}\Bigg|\sum_{(m,n)\in \mathrm{CT}}\braket{\nu|m,n}\Bigg|^2,
\end{equation}
where the sum is over CT site-pairs.

The constant recombination rate modified by the square of the sum of the amplitudes on relevant site-pairs agrees with generalised Marcus theory~\cite{Taylor2018} and with earlier recombination studies~\cite{Tempelaar2016}. In this work, we use $R_\mathrm{recomb}^\mathrm{xc}=\SI{e-11}{s^{-1}}$ and $R_\mathrm{recomb}^\mathrm{CT}=\SI{e-10}{s^{-1}}$.

\newpage

\section{dKMC algorithm}
We summarise the steps involved in the dKMC algorithm in \cref{alg:listing}.

\begin{algorithm*}[h]
    \normalsize
    \fbox{
    \begin{minipage}{0.95\textwidth}
        \setlist{nolistsep}
        \raggedright
        Given parameters $d$, $N$, $\sigma$, $\sigma_\mathrm{xc}$, $J_\mathrm{e}$, $J_\mathrm{h}$, $\mu$, $\lambda$, $\lambda_\mathrm{xc}$, $E_b$, $\omega_c$, $T$, $R_\mathrm{recomb}^\mathrm{xc}$, $R_\mathrm{recomb}^{CT}$, $a_\mathrm{dKMC}$, $r_\mathrm{sep}$, $d_\mathrm{xc}^\mathrm{max}$, $n_\mathrm{hops}$, $n_\mathrm{iter}$, and $n_\mathrm{traj}$:
        \begin{enumerate}[leftmargin=*]
            \item Calculate calibrating cutoff radii $r_\mathrm{hop}$ and $r_\mathrm{Ham}$ for charges and $r_\mathrm{hop}^\mathrm{xc}$ and $r_\mathrm{Ham}^\mathrm{xc}$ for excitons, using the procedure described in \cite{Balzer2023}.
            \item For $n_\mathrm{iter}$ realisations of disorder:
            \begin{enumerate}[leftmargin=*, label=\alph*.]
                \item Generate $N^d$ lattice of sites, with randomly oriented dipoles $\vec{\mu}$ and HOMO and LUMO energies drawn from the bivariate normal distributions in eq. (1) of the main text.
                \item For $n_\mathrm{traj}$ trajectories:
                \begin{enumerate}[leftmargin=*]
                    \item Choose the initial excitation location at a randomly chosen initial excitation distance $d_\mathrm{xc}\in [1,d_\mathrm{xc}^\mathrm{max}]$ from the interface.
                    \item Create a polaron-transformed $\tilde{H}_S$ containing all site-pairs within a combined distance of $r_\mathrm{Ham}$ of the chosen excitation location and all exciton site-pairs within $r_\mathrm{Ham}^\mathrm{xc}$. Diagonalise $\tilde{H}_S$ to find the polaron states and their energies, and calculate the expectation values of the positions of electron $\vec{C}^\mathrm{e}$, hole $\vec{C}^\mathrm{h}$, and exciton $\vec{C}^\mathrm{xc}$ in every state.
                    \item Create a list $L^\mathrm{xc}$ of exciton states (those with populations on exciton site-pairs greater than $p_\mathrm{cutoff}^\mathrm{xc}$) whose position is within 1 site of the excitation location.
                    \item Calculate the oscillator strength of each state in $L^\mathrm{xc}$ as the square of the expectation value of its transition dipole moment.
                    \item Choose the initial state $\nu$ from $L^\mathrm{xc}$ probabalistically in proportion to the states' oscillator strengths.
                    \item Set $n_\mathrm{sep}\leftarrow 0$.
                    \item For $n_\mathrm{hops}$ hops:
                    \begin{enumerate}[leftmargin=*]
                        \item If $\nu$ is an exciton state:
                        \begin{itemize}[leftmargin=*]
                            \item Diagonalise a new $\tilde{H}_S$, being a submatrix of the original $\tilde{H}_S$ describing site-pairs $(m,n)$ such that $\abs{\vec{r}_m-\vec{C}^\mathrm{e}_{\nu}} + \abs{\vec{r}_n-\vec{C}^\mathrm{h}_{\nu}} < r_\mathrm{Ham}$ and $(o,o)$ such that $\abs{\vec{r}_o-\vec{C}^\mathrm{xc}_{\nu}}< r_\mathrm{Ham}^\mathrm{xc}$.
                            \item Create a list $L$ containing states $\nu'$ such that $\abs{\vec{C}^\mathrm{e}_\nu-\vec{C}^\mathrm{e}_{\nu'}} + \abs{\vec{C}^\mathrm{h}_\nu-\vec{C}^\mathrm{h}_{\nu'}} < r_\mathrm{hop}$, and exciton states $\nu'$ such that $\abs{\vec{C}^\mathrm{xc}_\nu-\vec{C}^\mathrm{xc}_{\nu'}} < r^\mathrm{xc}_\mathrm{hop}$.
                        \end{itemize}
                        Else:
                        \begin{itemize}[leftmargin=*]
                            \item Diagonalise a new $\tilde{H}_S$, being a submatrix of the original $\tilde{H}_S$ describing site-pairs $(m,n)$ such that $\abs{\vec{r}_m-\vec{C}^\mathrm{e}_{\nu}} + \abs{\vec{r}_n-\vec{C}^\mathrm{h}_{\nu}} < r_\mathrm{Ham}$.
                            \item Create a list $L$ of all states $\nu'$ such that $\abs{\vec{C}^\mathrm{e}_\nu-\vec{C}^\mathrm{e}_{\nu'}} + \abs{\vec{C}^\mathrm{h}_\nu-\vec{C}^\mathrm{h}_{\nu'}} < r_\mathrm{hop}$.
                        \end{itemize}
                        \item Calculate $R_{\nu\nu'}$ for all $\nu'\in L$ using \cref{eq:secular_redfield_tensor}, only summing over each index in \cref{eq:Gamma} the minimum number of site-pairs $(m,n)$ that together support at least $a_\mathrm{dKMC}$ of the population of $\nu$ or $\nu'$.
                        \item Calculate $k^\nu_\mathrm{xc,recomb}$ using \cref{eq:xc_recomb} and append $g$ to $L$.
                        \item Calculate $k^\nu_\mathrm{CT,recomb}$ using \cref{eq:CT_recomb} and append $g$ to $L$.
                        \item Set $S_{\nu'}\leftarrow\sum_{\mu=1}^{\nu'} R_{\nu\mu}$ for all $\nu'\in L$ and set $T\leftarrow \sum_{\nu'\in L}S_{\nu'}$.
                        \item Find $\nu'$ such that $S_{\nu'-1} < uT < S_{\nu'}$, for uniform random number $u \in (0,1]$, and update $\nu\leftarrow\nu'$. 
                        \item Update $t\leftarrow t+\Delta t$, where $\Delta t = -T^{-1}\ln{v}$ for uniform random number $v \in (0,1]$.
                        \item If $\nu=g$, exit the for loop.
                        \item If $\abs{\vec{C}^\mathrm{e}_\nu-\vec{C}^\mathrm{h}_\nu}>r_\mathrm{sep}$, set $n_\mathrm{sep}\leftarrow n_\mathrm{sep}+1$ and exit the for loop.
                    \end{enumerate}
                \end{enumerate}
                 \item Calculate $\mathrm{IQE}=n_\mathrm{sep}/n_\mathrm{traj}$.
            \end{enumerate}
            \item Calculate mean IQE by averaging all IQEs.
        \end{enumerate}
    \end{minipage}
    }
    \caption{\textbf{dKMC for charge generation.}}
    \label{alg:listing}
\end{algorithm*}

\newpage

\section{Parameters}
\Cref{tab:parameters} contains the parameter values used in dKMC simulations throughout this work, unless otherwise stated.

\begin{table}[h]
	\centering
    \renewcommand{\arraystretch}{1.2}
    \normalsize
	\begin{tabular}{lll}
		\toprule
        Parameter & Description & Values \\
        \midrule
        $d$ & Dimension & 1--3\\
		$N$ & Sites along each dimension & 100\\
        $\sigma$ & Electronic disorder & 150 meV\\
        $\sigma_\mathrm{xc}$ & Exciton disorder & 30 meV\\
        $J_\mathrm{e}$ & Electron coupling & 7.5--75 meV\\
        $J_\mathrm{h}$ & Hole coupling & 7.5--75 meV\\
        $\mu$ & Transition dipole moment & 1--10 D\\
        $\lambda$ & Electronic bath reorganisation energy & 100 meV\\
        $\lambda_\mathrm{xc}$ & Exciton bath reorganisation energy & 100 meV\\
        $E_g$ & HOMO-LUMO gap & 1600 meV \\
        $E_b$ & Exciton binding energy & 700 meV\\
        $E^\mathrm{HOMO}_\mathrm{offset}$ & HOMO energetic offset & 500 meV\\
        $E^\mathrm{LUMO}_\mathrm{offset}$ & LUMO energetic offset & 500 meV \\
        $a$ & Lattice spacing & 1 nm \\
        $\varepsilon_r$ & Dielectric constant & 3.5 \\
        $\omega_c$ & Bath cutoff frequency & 62 meV\\ 
        $T$ & Temperature & 300 K \\
        $R_\mathrm{recomb}^\mathrm{xc}$ & Exciton recombination rate & $\num{e-11}~\mathrm{s}^{-1}$ \\
        $R_\mathrm{recomb}^\mathrm{CT}$ & CT recombination rate & $\num{e-10}~\mathrm{s}^{-1}$ \\
        $a_\mathrm{dKMC}$ & dKMC accuracy & 0.99 \\
        $r_\mathrm{sep}$ & Concluding separation & 5 nm\\ 
        $d_\mathrm{xc}^\mathrm{max}$ & Maximum excitation distance & 5 nm \\
        $n_\mathrm{hops}$ & Maximum hop number & 2000 \\
        $n_\mathrm{iter}$ & Simulation landscapes & 1000\\
        $n_\mathrm{traj}$ & Trajectories on each landscape &  10 in 2D, 1 in 3D\\
		\bottomrule
	\end{tabular}
	\caption{
	\textbf{Parameter values.} Default values used for parameters in dKMC simulations, unless otherwise specified.}
	\label{tab:parameters}
\end{table}

\newpage

\section{Properties of polaron states}
\begin{figure*}[h]
    \centering
    \includegraphics[width=\textwidth]{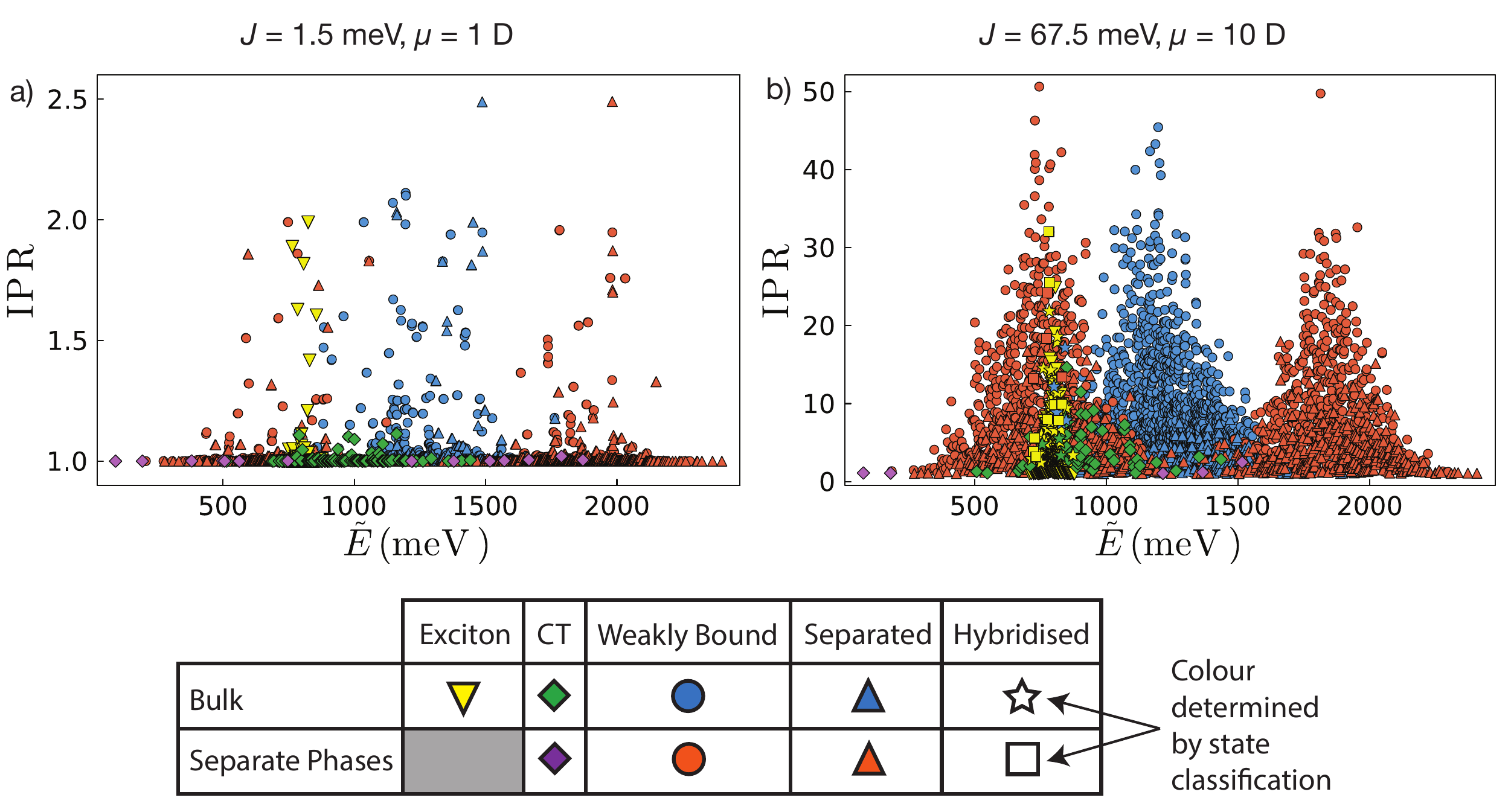}
    \caption{\textbf{Properties of the polaron states.} The inverse participation ratio (IPR) as a function of the energy $\tilde{E}$ of polaron states found by diagonalising a subset of the 2D polaron-transformed system Hamiltonian with
    \textbf{a)~}small electronic couplings and transition dipole moments ($J=\SI{1.5}{meV}$, $\mu=\SI{1}{D}$) and 
    \textbf{b)~}large ones ($J=\SI{67.5}{meV}$, $\mu=\SI{10}{D}$). Each state is labelled based on which category of site-pair it has the greatest overlap with and whether the site-pairs are in the bulk (same phase) or different phases.
    }
    \label{fig:states}
\end{figure*}

\Cref{fig:states} shows the relationship between the IPR and the energy $\tilde{E}$ of the states for  small couplings and dipole moments (\cref{fig:states}a) and larger ones (\cref{fig:states}b). \Cref{fig:states}a shows that when the couplings and dipole moments are small, the states are mostly localised onto one site-pair. For higher couplings and dipole moments, as in \cref{fig:states}b, the states can delocalised over more site-pairs. The IPR is greatest in the middle of the density of states, and smallest at the edges, where trap states tend to be localised. There are three peaks in the density of states due to the energetic offsets at the interface: the lowest-energy peak corresponds to the electron in the donor and the hole in the acceptor, the middle peak corresponds to the electron and hole in the same phase, and the highest-energy peak corresponds to the electron in the donor and the hole in the donor. Furthermore, as the states delocalise, they can form a significant number of hybridised states, seen as stars and squares in \cref{fig:states}b, while none form in the localised case of \cref{fig:states}a.

\newpage

\section{Mechanisms of delocalisation enhancements}
\begin{figure}[h]
    \centering
    \includegraphics[width=\textwidth]{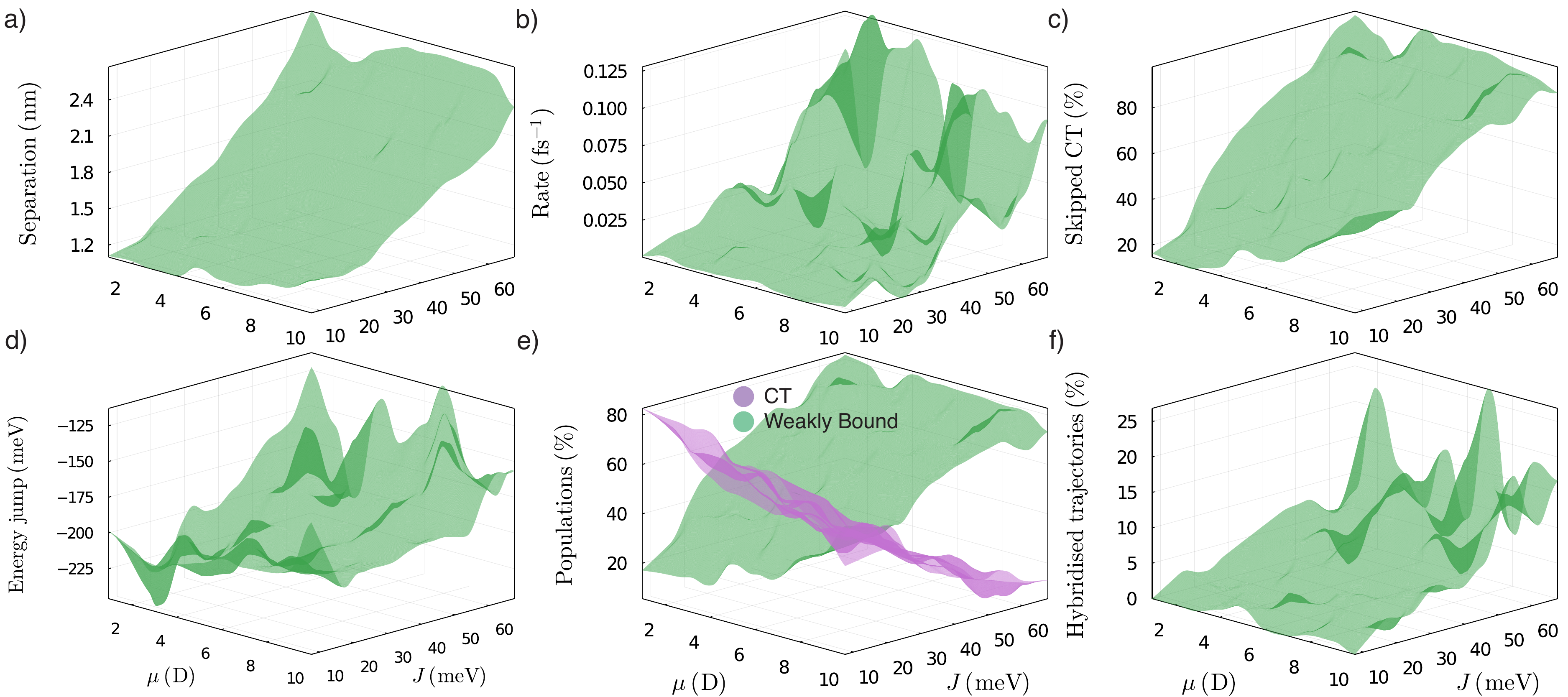}
    \caption{The same mechanistic analysis as in fig.~7 of the main text, but only for trajectories where the exciton dissociates across the interface.}
    \label{fig:mechanism_a}
\end{figure}

\begin{figure}[h]
    \centering
    \includegraphics[width=\textwidth]{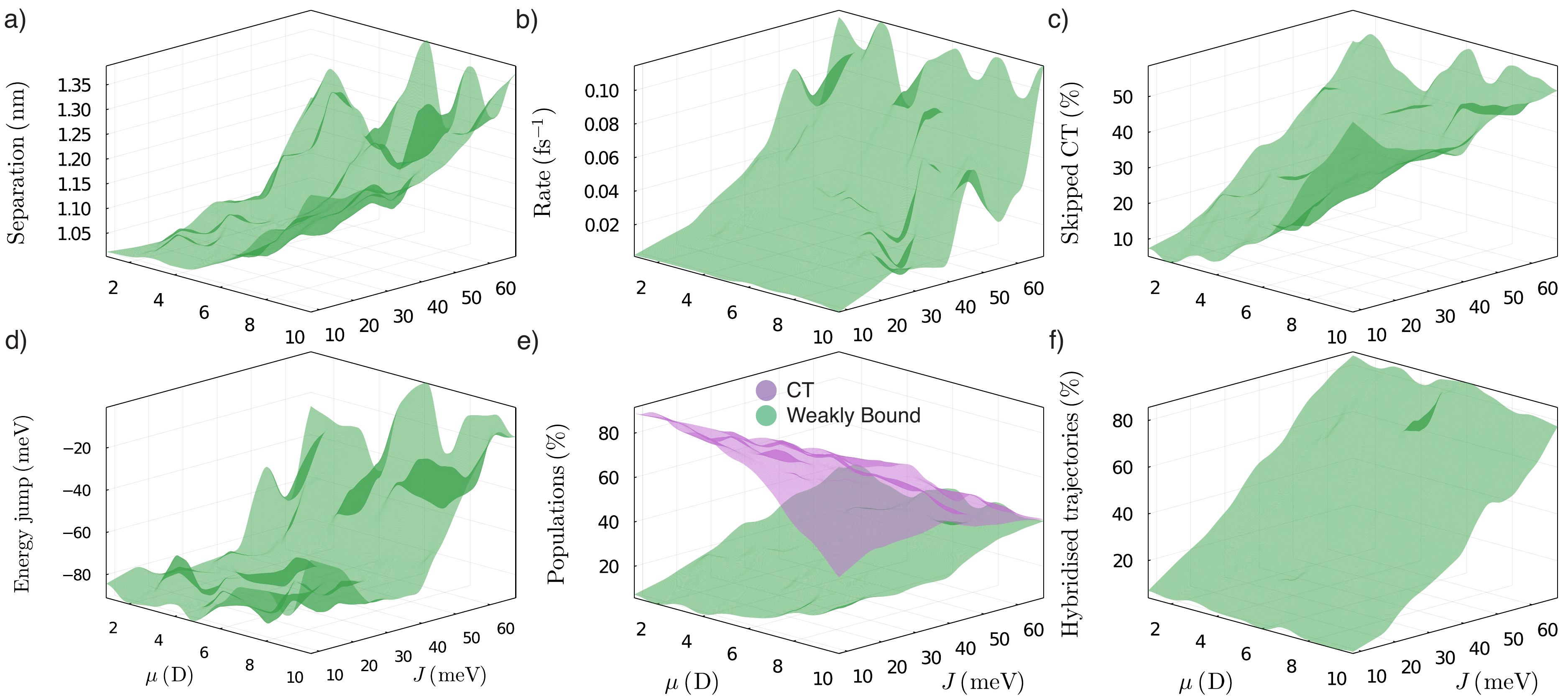}
    \caption{The same mechanistic analysis as in fig.~7 of the main text, but only for trajectories where the exciton dissociates in the bulk, without an interfacial energetic offset.}
    \label{fig:mechanism_b}
\end{figure}

Here, we break down the mechanistic analysis of section~II.D of the main text to show that the mechanisms of delocalisation enhancement are the same for both interfacial and bulk exciton dissociation.
To do so, instead of looking at all separated trajectories together, we separately analyse the trajectories where the exciton dissociates at the interface (\cref{fig:mechanism_a}) and those where it dissociates in the bulk, without an interfacial energetic offset (\cref{fig:mechanism_b}). In both cases, the same trends are seen as in fig.~7 of main text, i.e., that delocalisation both helps charges hop further out of an exciton and form hybridised states.

\newpage
\bibliography{bib}